\documentclass[journal]{IEEEtran}
\IEEEoverridecommandlockouts
 
\usepackage{amsmath}
\usepackage{cite}
\usepackage{amsmath,amssymb,amsfonts}
\usepackage{algorithm}
\usepackage{algpseudocode}
\usepackage{graphicx}
\usepackage{textcomp}
\usepackage{xcolor}
\usepackage{subcaption}
\usepackage{setspace}
\usepackage{booktabs}
\usepackage{array}
\usepackage{stfloats}
\usepackage{url}
\def\BibTeX{{\rm B\kern-.05em{\sc i\kern-.025em b}\kern-.08em
    T\kern-.1667em\lower.7ex\hbox{E}\kern-.125emX}}

\makeatletter

\newcommand{\Rmnum}[1]{\expandafter\@slowromancap\romannumeral #1@}
\makeatother

\makeatletter
\long\def\@makecaption#1#2{\ifx\@captype\@IEEEtablestring%
\footnotesize\begin{center}{\normalfont\footnotesize #1}\\
{\normalfont\footnotesize\scshape #2}\end{center}%
\@IEEEtablecaptionsepspace
\else
\@IEEEfigurecaptionsepspace
\setbox\@tempboxa\hbox{\normalfont\footnotesize {#1.}~~ #2}%
\ifdim \wd\@tempboxa >\hsize%
\setbox\@tempboxa\hbox{\normalfont\footnotesize {#1.}~~ }%
\parbox[t]{\hsize}{\normalfont\footnotesize \noindent\unhbox\@tempboxa#2}%
\else
\hbox to\hsize{\normalfont\footnotesize\hfil\box\@tempboxa\hfil}\fi\fi}
\makeatother

\makeatletter 
\pretocmd\@bibitem{\color{black}\csname keycolor#1\endcsname}{}{\fail}
\newcommand\citecolor[1]{\@namedef{keycolor#1}{\color{blue}}}
\makeatother
\citecolor{shi2018lstm}
\citecolor{rezende2015variational}
\citecolor{jiang2025electromagnetic}
\citecolor{zaheer2023survey}
\citecolor{tang2025low}
\citecolor{geraci2022will}

\begin{document}
\title{Conditional Generative Learning Enabled Wireless UAV Sensing and Tracking via Point Cloud Imaging}

\author{Xinhong~Dai,~\IEEEmembership{Graduate Student Member,~IEEE}, Yuan Gao, Hao Jiang,~\IEEEmembership{Graduate Student Member,~IEEE}, Xiaojun~Yuan,~\IEEEmembership{Fellow,~IEEE}, and Xin Wang,~\IEEEmembership{Fellow,~IEEE}
    \thanks{X. Dai, Y. Gao, and X. Wang are with Key Laboratory for Information Science of Electromagnetic Waves (MoE), College of Future Information Technology, Fudan University, Shanghai, China. (e-mail: \url{xhdai24@m.fudan.edu.cn}; \url{y_gao23@m.fudan.edu.cn}; \url{xwang11@fudan.edu.cn}).}
    \thanks{H. Jiang and X. Yuan are with the National Key Laboratory of Wireless Communications, the University of Electronic Science and Technology of China, Chengdu 611731, China (e-mail: \url{jh@std.uestc.edu.cn}; \url{xjyuan@uestc.edu.cn}).}
    \thanks{The corresponding author is
    \textit{Xin Wang}.}
    }
    
\maketitle
\begin{abstract}
In this paper, we study an unmanned aerial vehicle (UAV) sensing and tracking problem, where a base station equipped with an antenna array continuously illuminates a flying UAV and exploits the reflected echoes for slot-wise point cloud imaging within its potential flight region. To accomplish this task, the imaging region for each slot is determined based on the prior of the historical UAV positions. Then, the UAV is represented by an electromagnetic point cloud in this region that contains its spatial information and electromagnetic properties (EPs), enabling the unified extraction of UAV position, attitude, and shape from the reconstructed point cloud. The EP point cloud imaging for the UAV based on echo signals is a complex inverse problem. To this end, we propose an Array-based Unified Generative UAV Sensing and Tracking (AUGUST) approach, which integrates a conditional channel encoding module and a generative decoding module. The encoding module incorporates position and signal-to-noise ratio embeddings to stabilize the UAV intrinsic feature extraction under fast UAV position and channel variations, and maps the encoded features to a latent space regularized by a learnable flow-based prior. The decoding module employs a diffusion model with a weighted training objective to reconstruct the UAV point cloud guided by the extracted features. The simulation results demonstrate that the reconstructed point clouds via the proposed AUGUST approach present higher fidelity compared to the benchmark schemes, thereby enabling a more accurate capture of the UAV attitude and shape information. The AUGUST approach also presents a substantial gain over the conventional model-based baseline in positioning performance.
\end{abstract}


\IEEEpeerreviewmaketitle

\section{Introduction}
Benefiting from the flexible deployment, multifunctional capabilities, and other appealing features of unmanned aerial vehicles (UAVs), low-altitude economy (LAE) has attracted intense attention and is undergoing rapid development\cite{jiang2025integrated}. UAVs demonstrate their strong potential in different fields, including logistics, agriculture, and emergency communications\cite{toscano2024unmanned, Deng2024Radio,dai2023energy}, and have been recognized as an integral component of future sixth-generation wireless networks\cite{geraci2022will}. Looking ahead, the continuous expansion of the LAE is expected to lead to a substantial increase in the number of aerial vehicles\cite{fei2023air}. Consequently, accurately sensing and acquiring the flight states of UAVs is essential for ensuring low-altitude network security and robust operation. 

With the widespread adoption of integrated sensing and communication (ISAC) technologies in various wireless scenarios, the paradigm of wireless sensing for UAVs (especially echo-based non-cooperative sensing) based on antenna arrays deployed at base stations (BSs) has shown dominating superiority. This is attributed to the strong air–ground line-of-sight (LoS) of BS-UAV links, as well as the inherently superior information processing capability and broader sensing coverage provided by densely deployed cellular BSs\cite{geraci2022will}. In fact, array-based wireless sensing offers several unique advantages over conventional sensing modalities such as radar and LiDAR\cite{khawaja2025survey}. Antenna arrays serving the communication system provide wider sensing coverage in the three-dimensional (3D) space, particularly in the vertical domain, and are less sensitive to various environmental conditions. Also, the signal transmission of BS-UAV links can be directly exploited for sensing purposes, thereby avoiding extra wireless overhead. In addition, numerous elaborate waveform designs, along with advanced beamforming strategies, can be effectively applied to antenna arrays to enhance the sensing accuracy.

Many existing studies have investigated wireless non-cooperative sensing of UAVs using antenna arrays \cite{khawaja2025survey,fang2023jtea,fang20232,wen2023fast,dai2025attitude,you2023uav, Song2025cellular}. The works in \cite{fang2023jtea,fang20232,wen2023fast} focus predominantly on the estimation and tracking of a single attribute, such as UAV position. Notably, a flying UAV constitutes a rigid body with distinct attitude and geometric characteristics, and the evolving sensing requirements of LAE have further driven the shift toward joint sensing and tracking of both position and attitude \cite{dai2025attitude,you2023uav}, as well as the joint detection and classification of UAVs and other aerial targets \cite{khawaja2025survey, Song2025cellular}. However, the methodologies proposed in \cite{khawaja2025survey,fang2023jtea,fang20232,wen2023fast,dai2025attitude,you2023uav, Song2025cellular} are insufficient to meet the emerging demand for multidimensional UAV sensing, and a unified wireless sensing framework capable of comprehensively extracting such UAV attributes remains largely underdeveloped.

{\color{blue}From the perspective of electromagnetic propagation\cite{jiang2024paradigm}, characterizing or reconstructing these attributes from echo signals essentially forms a complex inverse problem, which maps received signals to an ``electromagnetic image'' of the UAV. Current target imaging and feature reconstruction methods fall into two broad categories: The first leverages radar imaging and compressed sensing-based techniques for target imaging and classification\cite{xu2022sparse}, while the second employs artificial intelligence (AI) algorithms to learn the deterministic mapping between received signals and target features\cite{xu2022sparse,zheng2023deep,guo2023physics}. The former typically relies on accurate statistical priors and explicit propagation modeling, which may be difficult to obtain in highly dynamic low-altitude scenarios. The latter can avoid explicit inverse solvers, but usually produces deterministic estimates and does not explicitly characterize the latent distribution or reconstruction uncertainty of the electromagnetic image conditioned on noisy echo observations. These limitations become particularly prominent for UAV sensing, where the echo-to-image mapping is high-dimensional, nonlinear, and potentially ill-posed due to limited observations, measurement noise, and rapid target-state variations.

In this context, generative AI (GAI) provides a distribution-aware perspective for wireless sensing inverse problems. Rather than only learning a deterministic mapping, generative models can capture the underlying distribution of physically plausible target representations and infer high-dimensional electromagnetic images from noisy and incomplete wireless observations through latent-variable modeling\cite{gilpin2024generative}. Such a capability is well aligned with UAV wireless sensing, where the received echoes contain implicit but ambiguous information about multidimensional target attributes.} For example, the study \cite{luo2021diffusion} presented a GAI-based point cloud generation method for 3D targets, where the generative model was used to transform the noise distribution to the distribution of a desired shape. Direct application of existing GAI-based methods to wireless imaging of highly dynamic UAVs, however, faces three core challenges: First, UAV mobility typically spans a large 3D space, rendering sensing and imaging across the entire flight domain computationally prohibitive; second, constructing a high-fidelity representation of UAV features over such an extensive spatial range poses a difficult task for algorithm design; third, measurement noise and rapid channel variations hinder the accurate extraction of intrinsic UAV features from echo signals.

Motivated by the above considerations, we investigate an array-based UAV sensing and tracking problem. Specifically, a BS equipped with an antenna array continuously transmits signals to a flying UAV and receives the corresponding echoes, based on which wireless imaging for the UAV is performed on a slot-by-slot basis. The signal transmission, reflection, and the sensing channel are modeled in accordance with an electromagnetic scattering framework. Note that UAV motion is not entirely random; instead, its trajectory exhibits temporal continuity, with strong correlations between consecutive positions. This continuity enables historical position information to serve as valuable spatial prior knowledge during tracking, obviating the need for full-flight-region sensing. Hence, by exploiting the prior of historical UAV positions, the UAV position for each slot can be predicted, thereby constraining a potential flight region for wireless imaging centered at the predicted position. Then, we adopt the electromagnetic point cloud to visualize the sensed UAV within this region, where each point contains the spatial information and electromagnetic properties (EPs) of each scattering point on the UAV. By following the GAI principle, we propose an Array-based Unified Generative UAV Sensing and Tracking (AUGUST) approach. This approach not only can efficiently extract UAV features from array-received echo signals but also integrate these features with spatial prior information under a Bayesian framework, thereby achieving efficient and high-quality EP point cloud-based UAV imaging. The main contributions of this paper are summarized as follows:
\begin{itemize}
 \item We formulate a UAV sensing and tracking problem that aims to reconstruct the EP point cloud of the UAV within the predicted flight region, conditioned on the estimated sensing channel. To solve this problem, we propose an AUGUST approach comprising a conditional channel encoding module and a generative decoding (reconstruction) module, to capture the UAV position, attitude, and shape information from the visualized point cloud.
 \item In the conditional channel encoding module, a multiplicative position embedding and a signal-to-noise ratio (SNR) embedding are incorporated to help the encoder, structured with a multilayer perceptron (MLP), effectively extract intrinsic UAV features with changing UAV locations and channel conditions. Then, the encoded features are mapped into a latent space regularized by a learnable global prior, to improve the generalization and consistency of the model.
 \item Given the extracted features by the conditional channel encoding, the generative decoding module exploits the diffusion model to generate the UAV point cloud. To mitigate the adverse impact of distribution discrepancies among different features on the training, we develop a weighted training design by introducing a weighted loss that accounts for the UAV spatial information, EPs, and the regularization prior.
\end{itemize}

Numerical results show that the proposed AUGUST approach achieves higher-quality point cloud imaging compared to the alternatives, resulting in more accurate UAV attitude and shape characterization. Also, the AUGUST approach outperforms the conventional positioning method in position estimation performance.
\subsection{Related Works}
Recently, GAI has demonstrated immense potential in semantic transmission \cite{jiang2025blind}, channel representation \cite{yue2023dif}, and ISAC systems \cite{chen2025scoring}. Beyond these domains, several representative GAI models—such as diffusion models—exhibit high flexibility in both modeling and inference design, enabling their effective adaptation to wireless sensing scenarios. Specifically, the authors in \cite{wang2024generative} employed a weighted conditional diffusion model to achieve human flow detection. In \cite{wang2025human}, a diffusion-based framework was developed to generate high-quality radio-frequency synthetic data for human pose sensing. Additionally, the authors in \cite{jiang2024electromagnetic} modeled the EP point cloud of targets based on electromagnetic scattering principles and leveraged a diffusion model to realize 3D point cloud imaging, with the sensing channel serving as a conditioning factor. Here, the channel was mapped to a reference channel to guide the reverse diffusion process.

As a comparable EP point cloud imaging approach based on echo signals, \cite{jiang2024electromagnetic} focuses on static target sensing but the methodology can still be used as the point cloud imaging component in our proposed AUGUST approach, once the point-cloud reconstruction region is specified. Different from \cite{jiang2024electromagnetic}, we do not directly feed the channel information into the diffusion model. Instead, we adopt channel encoding with position and SNR embeddings, coupled with latent space mapping under the variational autoencoder (VAE) framework \cite{sohn2015learning}. {\color{blue}More recently, a diffusion Schrödinger bridge (DSB)-based framework was proposed to bridge EM-property sensing and channel reconstruction in ISAC systems\cite{jiang2025electromagnetic}, which further demonstrates the potential of integrating electromagnetic scattering knowledge with generative reconstruction methods. Different from this bidirectional EM-property/channel reconstruction framework, this paper focuses on dynamic UAV sensing and tracking via slot-wise EP point cloud imaging, where the reconstructed point cloud is further used to infer the UAV position, attitude, and shape.}

\subsection{Organization and Notation}
\par The rest of this paper has the following structure: In Section \Rmnum{2}, we introduce the signal model and the sensing model under the electromagnetic scattering framework. In Section \Rmnum{3}, we formulate a UAV sensing and tracking problem where the objective is UAV point cloud reconstruction. In Section \Rmnum{4}, we develop the AUGUST approach to solve the problem based on the GAI principle. In Section \Rmnum{5}, we introduce a weighted training design for the proposed framework. Numerical results are provided in Section \Rmnum{6}, followed by the conclusion in Section \Rmnum{7}.

\textit{Notations:} Bold lowercase and uppercase letters denote vectors and matrices, respectively. $\ell^2$ norm and Frobenius norm are denoted by $\|\cdot\|$ and $\|\cdot\|_F$, respectively. The transpose, conjugate transpose, and inverse of a matrix are represented by $(\cdot)^{\mathsf{T}}$, $(\cdot)^{\mathsf{H}}$, and $(\cdot)^{-1}$, respectively. $\text{diag}[\cdot]$ means forming a diagonal matrix with the given elements placed on the diagonal. $\text{vec}(\boldsymbol{A})$ and $\text{tr}(\boldsymbol{A})$ denote the vectorization and trace of matrix $\boldsymbol{A}$, respectively. $D_{\text{KL}}(p(\cdot)\|q(\cdot))$ denotes the Kullback–Leibler (KL) divergence from distribution $p(\cdot)$ to distribution $q(\cdot)$. $\otimes$, $\odot$, and $\nabla$ denote the Kronecker product, Hadamard product, and gradient operator, respectively. $\mathcal{CN}$ and $\mathcal{N}$ represent the circularly symmetric complex Gaussian (CSCG) distribution and the real-valued Gaussian distribution, respectively. $\mathbf{I}_N$ denotes the identity matrix of size $N$. $\text{j}$ is the imaginary unit such that $\text{j}^2 = -1$. The expectation of $\boldsymbol{x}$ with respect to the distribution $p$ is denoted by $\mathbb{E}_p[\boldsymbol{x}]$. The index set $\mathcal{I}_{N}$ is defined as $\{1,\ldots, N\}$ for integer $N$.

 \section{System Model}
 \subsection{Signal Model}
\begin{figure}[t]
    \centering   \includegraphics[width=.98\linewidth]{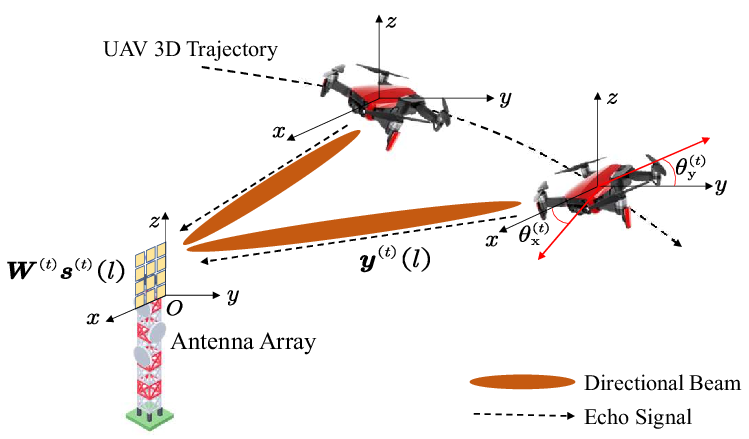}
    \caption{The considered UAV sensing and tracking system.}
    \label{system model}
\end{figure}

As illustrated in Fig. \ref{system model}, we consider a UAV sensing and tracking system, in which the BS transmits signals to a UAV flying in 3D space and exploits the reflected echoes for UAV sensing and tracking. In this system, the BS is equipped with an antenna array deployed in the $x$-$O$-$z$ plane, consisting of $N_b = N_{\mathsf{x}} \times N_{\mathsf{z}}$ antennas for transmission and reception. All antennas are uniformly spaced at half of the carrier wavelength $\lambda_{\rm c} = \frac{c}{f_c}$, where $c$ denotes the speed of light and $f_c$ is the carrier frequency. In this paper, we only consider the LoS link between the UAV and the BS for analytical tractability. Moreover, we assume that the UAV sensing and tracking operate over $T$ discrete time slots indexed by $t$, and each slot has an equal length $\Delta_t$. 

Since the BS cannot accurately obtain the UAV position during tracking, a rough beamforming design, over a potential region within which the UAV is flying, is adopted based on a discrete Fourier transform (DFT) codebook\cite{han2018dft}. This design enables uniform beam coverage over this region of interest (RoI) while striking a favorable trade-off between beam scanning range and echo power. Specifically, let $\vartheta$ and $\varphi$ denote the azimuth and elevation angles, respectively. The RoI is modeled as a 2D angular sector centered at direction $(\vartheta^{(t)}_{\text{c}}, \varphi^{(t)}_{\text{c}})$ with angular widths $\Delta\vartheta$ and $\Delta\varphi$. Define the uniform angle grids for the DFT codebook as $\{\vartheta^{(t)}_\imath\}_{\imath=1}^
{N_{\mathsf{x}}}\in[\vartheta^{(t)}_{\text{c}}-\frac{\Delta\vartheta}{2},\vartheta^{(t)}_{\text{c}}+\frac{\Delta\vartheta}{2}]$ and $\{\varphi^{(t)}_\jmath\}_{\jmath=1}^
{N_{\mathsf{z}}}\in[\varphi^{(t)}_{\text{c}}-\frac{\Delta\varphi}{2},\varphi^{(t)}_{\text{c}}+\frac{\Delta\varphi}{2}]$. For each slot $t$, we consider that the BS transmits $L$ symbols, and the $l$-th transmitted signal in vector form is given by:
\begin{align}
    \boldsymbol{x}^{(t)}(l) = \boldsymbol{W}^{(t)}\boldsymbol{s}^{(t)}(l)\in \mathbb{C}^{N_b\times 1},\quad l\in \mathcal{I}_L,
    \label{signal x}
\end{align}
where $\boldsymbol{W}^{(t)}=[\boldsymbol{w}^{(t)}_{1,1},\cdots,\boldsymbol{w}^{(t)}_{\imath,\jmath},\cdots,\boldsymbol{w}^{(t)}_{N_\mathsf{x},N_\mathsf{z}}]\in \mathbb{C}^{N_b \times N_b}$, $\imath\in \mathcal{I}_{N_\mathsf{x}}, \jmath\in \mathcal{I}_{N_\mathsf{z}}$, and $\boldsymbol{s}^{(t)}(l)\in \mathbb{C}^{N_b \times 1}$ denotes the $l$-th downlink symbol vector intended to the UAV. The $(\imath,\jmath)$-th column of  $\boldsymbol{W}^{(t)}$, i.e., $\boldsymbol{w}^{(t)}_{\imath,\jmath}$ is an $N_b$-dimensional vector, and its $n$-th element is given by
\begin{align}
\label{beamforming matrix element}
[\boldsymbol{w}^{(t)}_{\imath,\jmath}]_{n} = \frac{1}{\sqrt{N_b}} e^{-\text{j} \pi (a\cos{\varphi^{(t)}_{\jmath}}\cos{\vartheta^{(t)}_{\imath}}+b\sin{\varphi^{(t)}_{\jmath}})},
\end{align}
with $a\in \mathcal{I}_{N_\mathsf{x}}, b\in \mathcal{I}_{N_\mathsf{z}}$, and $n = b +(a-1)N_{\mathsf{z}}\in \mathcal{I}_{N_{b}}$. In addition, the transmitted symbols are modeled as pseudo-random, satisfying $\mathbb{E}[\boldsymbol{s}^{(t)}(l)] = \mathbf{0}$ and $\mathbb{E}[\boldsymbol{s}^{(t)}(l)(\boldsymbol{s}^{(t)}(l))^\mathsf{H}] = \mathbf{I}_{N_b}$. Then, let $\boldsymbol{X}^{(t)} = [\boldsymbol{x}^{(t)}(1),\cdots,\boldsymbol{x}^{(t)}(L)]\in \mathbb{C}^{N_b \times L}$ denote the transmitted signal matrix over $L$ symbols. When the number of symbols $L$ is sufficiently large, the sample covariance matrix of $\boldsymbol{X}^{(t)}$ can be approximated as $\boldsymbol{S}^{(t)}_{\boldsymbol{X}}\approx \frac{1}{L}\boldsymbol{X}^{(t)}(\boldsymbol{X}^{(t)})^\mathsf{H}$. Besides, we consider that the BS transmits signals subject to a fixed power constraint $P_{\text{s}}$, i.e., $\text{tr}(\boldsymbol{S}^{(t)}_{\boldsymbol{X}})\le P_{\text{s}}$.
\subsection{Sensing Model Based on Electromagnetic Scattering}
For each slot $t$, we assume that the antenna array at the BS operates in full-duplex mode, and the received signals at the BS can be modeled as 
\begin{align}
    \boldsymbol{Y}^{(t)} = \boldsymbol{H}^{(t)}\boldsymbol{X}^{(t)}+\boldsymbol{N}.
    \label{received echo signal}
\end{align}
In \eqref{received echo signal}, $\boldsymbol{Y}^{(t)}=[\boldsymbol{y}^{(t)}(1),\boldsymbol{y}^{(t)}(2),\cdots,\boldsymbol{y}^{(t)}(L)]$ is the received signal matrix, and $\boldsymbol{N}=[\boldsymbol{n}(1),\boldsymbol{n}(2),\cdots,\boldsymbol{n}(L)]$ represents the CSCG noise with the variance $\sigma^2$ at the BS, i.e., $\boldsymbol{n}(l)\sim \mathcal{CN}(\boldsymbol{0},\sigma^2 \mathbf{I}_{N_b})$. $\boldsymbol{H}^{(t)}\in \mathbb{C}^{N_b \times N_b}$ denotes the sensing channel matrix corresponding to the downlink transmission and signal reflection processes. We next detail the sensing model of the considered system based on the electromagnetic scattering modeling. 
Note that the flying UAV is a moving scattering medium with a specific geometry and electromagnetic characteristics. The EPs of the UAV are commonly manifested as spatial distributions of relative permittivity and conductivity, which capture the EP contrast between the UAV and the surrounding air medium. Since the relative permittivity and conductivity of the air are approximately equal to 1 and 0, respectively, the distributions can be formulated via the contrast function \cite{liu2019subspace} defined as
\begin{align}
     \label{contrast function}
    \chi(\boldsymbol{p}^{(t)}) = (\varepsilon-1)+\frac{\text{j}\varrho}{2\pi f_c\varepsilon_0},\quad \boldsymbol{p}^{(t)}\in \mathbb{R}^{3\times 1},
\end{align}
where $\varepsilon$ and $\varrho$ represent the relative permittivity and conductivity at point $\boldsymbol{p}^{(t)}$ on the UAV, respectively. In this paper, we assume that the scatter points on the UAV share the same EPs. Besides, since the EPs of the UAV remain invariant during the tracking, the dependence on $t$ and $\boldsymbol{p}^{(t)}$ of both $\varepsilon$ and $\varrho$ is omitted for notational simplicity. $\varepsilon_0$ is the vacuum permittivity.
{\color{blue}Note that the UAV can be regarded as quasi-static during one slot-wise channel estimation and imaging operation. For the considered single-carrier pilot setting with a moderate bandwidth, the pilot duration of a slot-wise sensing is typically on the order of microseconds. In this case, the corresponding displacement, Doppler-induced phase variation, and attitude variation of the UAV within such a short duration are negligible.}

Each antenna of the BS array can be modeled as an excitation current source, and the currents on the transmitting antennas linearly induce the electric field. Hence, for the $l$-th transmitted signal, the incident electric field at $\boldsymbol{p}^{(t)}$ generated by the whole array can be expressed as\cite{bassen2003electric}
\begin{align}
    \boldsymbol{E}^{(t)}_{\mathsf{i}}(\boldsymbol{p}^{(t)},l)&=[\boldsymbol{m}^{(t)}_{1,1}(\boldsymbol{p}^{(t)}),\cdots,\boldsymbol{m}^{(t)}_{N_{\mathsf{x}},N_{\mathsf{z}}}(\boldsymbol{p}^{(t)})]\boldsymbol{x}^{(t)}(l),\nonumber\\  &=\boldsymbol{M}^{(t)}(\boldsymbol{p}^{(t)})\boldsymbol{x}^{(t)}(l).
    \label{Ei}
\end{align}
The matrix $\boldsymbol{M}^{(t)}(\boldsymbol{p}^{(t)}) \in \mathbb{C}^{3\times N_b}$ maps the signal excitation $\boldsymbol{x}^{(t)}(l)$ to the incident electric field $\boldsymbol{E}^{(t)}_{\mathsf{i}}(\boldsymbol{p}^{(t)}, l)$. The vector $\boldsymbol{m}^{(t)}_{\imath,\jmath}(\boldsymbol{p}^{(t)})$ denotes the electric field at point $\boldsymbol{p}^{(t)}$ generated by the $(\imath,\jmath)$-th antenna under unit current excitation. Accordingly, $\boldsymbol{E}^{(t)}_{\mathsf{i}}(\boldsymbol{p}^{(t)}, l)$ can be interpreted as a linear superposition of the electric fields generated by all array antennas. Specifically, $\boldsymbol{m}^{(t)}_{\imath,\jmath}(\boldsymbol{p}^{(t)})$ is defined as 
\begin{align}
    \label{Ei n}
    \boldsymbol{m}^{(t)}_{\imath,\jmath}(\boldsymbol{p}^{(t)}) = \int \overline{\overline{\boldsymbol{G}}}(\boldsymbol{p}^{(t)},\boldsymbol{p}'_{\imath,\jmath})\overline{\boldsymbol{J}}(\boldsymbol{p}'_{\imath,\jmath})\text{d}\boldsymbol{p}'_{\imath,\jmath},
\end{align}
{\color{blue}where $\overline{\boldsymbol{J}}(\boldsymbol{p}'_{\imath,\jmath})$ denotes the equivalent current distribution induced at point $\boldsymbol{p}'_{\imath,\jmath}$ on the $(\imath,\jmath)$-th antenna under unit current excitation.} In this paper, we simplify the calculation of the integration in \eqref{Ei n} by approximately modeling each antenna as a point dipole located at $\boldsymbol{p}_{\imath,\jmath}$. $ \overline{\overline{\boldsymbol{G}}}(\boldsymbol{p}^{(t)},\boldsymbol{p}'_{\imath,\jmath})\in \mathbb{C}^{3\times 3}$ in \eqref{Ei n} is the dyadic electric field Green’s function\cite{li1994electromagnetic}. Given any two points $\boldsymbol{p}_1$ and $\boldsymbol{p}_2$, $\overline{\overline{\boldsymbol{G}}}(\boldsymbol{p}_1, \boldsymbol{p}_2)$ can be specifically formulated as 
\begin{align}
\label{green funcion}
\overline{\overline{\boldsymbol{G}}}
(\boldsymbol{p}_1, \boldsymbol{p}_2)
= &\left(\mathbf{I}_3+\frac{\nabla\nabla}{k_c^2}\right)\frac{e^{\text{j} k_c r_{12}}}{4\pi r_{12}}\nonumber\\
=&\left[(g(r_{12})\overline{\boldsymbol{r}}_{12}\overline{\boldsymbol{r}}^{\mathsf{T}}_{12}-h(r_{12})\mathbf{I}_3\right]\frac{e^{\text{j} k_c r_{12}}}{4\pi r_{12}},
\end{align}
where $k_c = \frac{2\pi}{\lambda_c}$ is the wave number in the air, $r_{12}=\|\boldsymbol{p}_1 - \boldsymbol{p}_2\|$, $\overline{\boldsymbol{r}}_{12}=\frac{\boldsymbol{p}_1 - \boldsymbol{p}_2}{\|\boldsymbol{p}_1 - \boldsymbol{p}_2\|}$, $g(r_{12})=\frac{3}{k^2_c r^2_{12}}-\frac{3\text{j}}{k_c r_{12}}-1$, and $h(r_{12})=\frac{1}{k^2_c r^2_{12}}-\frac{\text{j}}{k_c r_{12}}-1$. When the incident electric field $\boldsymbol{E}^{(t)}_{\mathsf{i}}(\boldsymbol{p}^{(t)}, l)$ illuminates the UAV along the beam direction, it induces conduction currents on the UAV surface. The UAV then acts as an equivalent current source and re-radiates to produce the total electric field $\boldsymbol{E}^{(t)}_{\mathsf{tot}}(\boldsymbol{p}^{(t)}, l)$. As a result, according to the Lippmann-Schwinger equation\cite{chen2018computational}, for a 3D region $\boldsymbol{p}'\in \mathcal{R}_{\mathsf{u}}$ in which the flying UAV is located, $\boldsymbol{E}^{(t)}_{\mathsf{tot}}(\boldsymbol{p}^{(t)},l)$ within $\mathcal{R}_{\mathsf{u}}$ can be obtained as 
\begin{align}
    \label{LS eq}
   \boldsymbol{E}^{(t)}_{\mathsf{tot}}(\boldsymbol{p}^{(t)},l) = &k^2_c\iiint_{\mathcal{R}_{\mathsf{u}}}\chi(\boldsymbol{p}')\overline{\overline{\boldsymbol{G}}}(\boldsymbol{p}^{(t)},\boldsymbol{p}')\boldsymbol{E}^{(t)}_{\mathsf{tot}}(\boldsymbol{p}',l)\text{d}\boldsymbol{p}'\nonumber\\
    &+\boldsymbol{E}^{(t)}_{\mathsf{i}}(\boldsymbol{p}^{(t)},l).
\end{align}
Then, similar to \eqref{LS eq}, the scattered electric field $\boldsymbol{E}^{(t)}_{\mathsf{s}}(\boldsymbol{p}_{\imath,\jmath},l)$ back to the $(\imath,\jmath)$-th receiving antenna is given by\cite{vargas2022subspace} 
\begin{align}
    \label{wave eq Es}
    \boldsymbol{E}^{(t)}_{\mathsf{s}}(\boldsymbol{p}_{\imath,\jmath},l) = &k^2_c\iiint_{\mathcal{R}_{\mathsf{u}}}\chi(\boldsymbol{p}')\overline{\overline{\boldsymbol{G}}}(\boldsymbol{p}_{\imath,\jmath},\boldsymbol{p}')\boldsymbol{E}^{(t)}_{\mathsf{tot}}(\boldsymbol{p}',l)\text{d}\boldsymbol{p}'.
\end{align}
As the inverse process of \eqref{Ei}, the scattered electric field $\boldsymbol{E}^{(t)}_{\mathsf{s}}(\boldsymbol{p}_{\imath,\jmath},l)$ will be converted into the electrical signal at each antenna. Thus, the $l$-th received signal at the BS with the unit receiving antenna gain is given by 
\begin{align}
    \label{signal yl}
    \boldsymbol{y}^{(t)}(l) = [\boldsymbol{E}^{(t)}_{\mathsf{s}}(\boldsymbol{p}_{1,1},l),\cdots,\boldsymbol{E}^{(t)}_{\mathsf{s}}(\boldsymbol{p}_{N_{\mathsf{x}},N_{\mathsf{z}}},l)]^\mathsf{T}\overline{\boldsymbol{v}}_p+\boldsymbol{n}(l),
\end{align}
where the unit vector $\overline{\boldsymbol{v}}_p\in \mathbb{R}^{3\times 1}$ denotes the polarization vector of the receiving array. {\color{blue}Note that the above electromagnetic sensing model differs from conventional acoustic-, optical-, and radar-based UAV sensing paradigms\cite{zaheer2023survey,tang2025low,khawaja2025survey}, which rely on the particular target feature forms and signal design. The equations from \eqref{Ei} to \eqref{signal yl} unfold the electromagnetic propagation mechanism in \eqref{received echo signal}. This indicates that the spatial and EP information of the UAV can be implicitly embedded into the single-carrier array sensing channel through vector electromagnetic propagation, EP-dependent scattering, and polarization projection at the receiving antennas, thereby providing a physical basis for structure-level UAV sensing that is independent of multicarrier transmission and delay estimation.} In this case, $\boldsymbol{H}^{(t)}$ can be utilized as prior knowledge to reconstruct the distribution of $\chi(\boldsymbol{p}^{(t)})$. To this end, the least squares (LS) method \cite{biguesh2006training} is adopted to estimate the sensing channel, denoted as $\boldsymbol{H}^{(t)}_{\mathsf{est}}$. The estimated $\boldsymbol{H}^{(t)}_{\mathsf{est}}$ is then employed as prior information to successively reconstruct the distribution of $\chi(\boldsymbol{p}^{(t)})$ in a slot-wise manner, thereby enabling the unified sensing and tracking of the UAV’s position, attitude, and shape.

\section{Problem Formulation}
To effectively capture the multidimensional information of the UAV, it is necessary to reconstruct and image its EP distribution in the 3D space. Traditional reconstruction methods for electromagnetic imaging, such as pixel-based methods\cite{zhang2022probabilistic}, typically require analysis of the entire RoI, including the background medium. This incurs additional computational burden and is not suitable for the considered real-time UAV tracking scenario. Furthermore, our focus is on reconstructing information about the UAV itself. Therefore, a point cloud-based representation is adopted to visualize the EP distribution of the UAV, enabling an efficient, clear, and direct representation of its multidimensional information by separating the target from the background medium \cite{luo2021diffusion}.

As discussed earlier, both the beamforming design and point cloud reconstruction rely on a predefined position to determine the beam direction and the reconstruction region within a limited spatial region. As a result, for each slot $t$ during the tracking, we aim to utilize the temporal correlation of historical UAV locations to predict the UAV position (denoted as $\boldsymbol{q}_{\mathsf{pre}}^{(t)} = [\tilde{q}^{(t)}_x,\tilde{q}^{(t)}_y,\tilde{q}^{(t)}_z]^{\mathsf{T}}$).
Then, the point cloud reconstruction for slot $t$ can be performed within a certain RoI centered on $\boldsymbol{q}^{(t)}_{\mathsf{pre}}$ and bounded by a prediction error threshold.
Based on these considerations, we represent the UAV as a 5D point cloud at slot $t$, which is composed of $M$ normalized points, i.e., 
$\mathcal{P}^{(t)} = \{\boldsymbol{p}^{(t)}_i |i\in \mathcal{I}_M\}$. Each 5D point in $\mathcal{P}^{(t)}$ contains two parts, including the 3D coordinate information and 2D EPs information, which is given by 
\begin{align}
    \label{5D pc}
    \boldsymbol{p}^{(t)}_i = \left[\frac{x^{(t)}_i-\tilde{q}^{(t)}_x}{s_x},\frac{y^{(t)}_i-\tilde{q}^{(t)}_y}{s_y},\frac{z^{(t)}_i-\tilde{q}^{(t)}_z}{s_z},\frac{\varepsilon}{\varepsilon_{\mathsf{max}}},\frac{\varrho}{\varrho_{\mathsf{max}}}\right]^{\mathsf{T}},
\end{align}
where $x^{(t)}_i$, $y^{(t)}_i$, and $z^{(t)}_i$ represent the coordinates of the $i$-th point along the corresponding dimension; $\varepsilon_{\mathsf{max}}$ and $\varrho_{\mathsf{max}}$ respectively denote the maximum relative permittivity and conductivity based on the electromagnetic material settings in the dataset, as detailed in the following. {\color{blue}Here, $s_x$, $s_y$, and $s_z$ denote the standard deviations of the reconstruction region, which are determined by the UAV size and the position prediction error thresholds in the corresponding dimensions. In general, a smaller $s_{x/y/z}$ indicates a narrower predicted flight region and thus provides a stronger spatial prior for UAV point cloud reconstruction. However, for highly dynamic trajectories with larger speed variations, more frequent velocity-direction changes, or stronger acceleration variations, the prediction error of $\boldsymbol{q}^{(t)}_{\mathsf{pre}}$ may increase, and a larger $s_{x/y/z}$ is required to ensure that the UAV is covered by the reconstruction region. It is worth mentioning that $s_{x/y/z}$ can be empirically set in advance by assessing the randomness of UAV flight or the trajectory prediction error before the tracking and imaging, introducing a trade-off between robustness to trajectory prediction errors and point cloud reconstruction quality.}

Note that $\mathcal{P}^{(t)}$ provides a discrete representation of the intrinsic physical properties of the UAV and can be regarded as a collection of samples drawn from an underlying high-dimensional distribution $p(\mathcal{P}^{(t)})$ that characterizes these properties. Furthermore, $\boldsymbol{H}^{(t)}_\mathsf{est}$ implicitly incorporates the multidimensional information of the UAV, which can be utilized as prior information to facilitate the point cloud reconstruction. Therefore, the UAV sensing and tracking problem can be modeled as the maximization of the conditional probability given the measurement information and historical data, i.e., 
\begin{align}
     \label{problem formulation 1}
    \arg \max_{\boldsymbol{p}^{(t)}_i} \quad p(\mathcal{P}^{(t)}|\boldsymbol{H}^{(t)}_\mathsf{est},\hat{\mathcal{P}}^{(1:t-1)}), \quad \forall t\in \mathcal{I}_{T}.
\end{align}
In \eqref{problem formulation 1}, $\hat{\mathcal{P}}^{(1:t-1)}$ denote the historical reconstructed point clouds before slot $t$, which are used for the prediction of $\boldsymbol{q}^{(t)}_{\mathsf{pre}}$. Specifically, the UAV positions can be estimated via the geometric centroid of all point cloud coordinates, i.e.,
$\boldsymbol{q}^{(1:t-1)}_{\mathsf{est}} = \frac{1}{M}\sum_{i=1}^{M}\hat{\boldsymbol{p}}^{(1:t-1)}_{\text{pos};i}$, where $\hat{\boldsymbol{p}}^{(1:t-1)}_{\text{pos};i}$ denote the 3D coordinates of the $i$-th reconstructed 5D points. {\color{blue}In implementation, the velocity and acceleration information can be derived from the historical trajectory and used to form a kinematic prediction $\boldsymbol{q}^{(t)}_{\mathsf{kin}}$. Since practical UAV trajectories may contain nonlinear maneuvers and abrupt direction changes, an LSTM-based residual predictor can be further employed to learn a correction term $\Delta\boldsymbol{q}^{(t)}_{\mathsf{LSTM}}$ from historical position, velocity, and acceleration features \cite{shi2018lstm}. The final predicted center can be obtained via $\boldsymbol{q}^{(t)}_{\mathsf{pre}}=\boldsymbol{q}^{(t)}_{\mathsf{kin}}+\Delta\boldsymbol{q}^{(t)}_{\mathsf{LSTM}}$.}

Given $\boldsymbol{q}^{(t)}_{\mathsf{pre}}$, problem \eqref{problem formulation 1} is recast as a point cloud reconstruction problem within a deterministic region conditioned on the channel data, i.e., maximizing $p(\mathcal{P}^{(t)}|\boldsymbol{H}^{(t)}_\mathsf{est})$ for each slot $t$. In general, maximum a posteriori (MAP) or minimum mean square error (MMSE) estimation can be employed to solve this problem. However, it remains challenging to explicitly characterize both $p(\mathcal{P}^{(t)})$ and $p(\mathcal{P}^{(t)}|\boldsymbol{H}^{(t)}_\mathsf{est})$, owing to the intricate mapping from the multidimensional information of the UAV to the channel matrix. To address this difficulty, we next leverage the powerful data distribution learning capabilities of generative learning and propose an efficient AUGUST approach by following the GAI principle. This approach can effectively capture the underlying high-dimensional features embedded in the measurement data, enabling more accurate and data-driven point cloud reconstruction.

\section{Proposed AUGUST Approach}
Since the conditional probability  $p(\mathcal{P}^{(t)}|\boldsymbol{H}^{(t)}_\mathsf{est})$ is difficult to obtain directly, we address this issue from a generative learning perspective by training a parameterized probabilistic model $p_{\boldsymbol{\theta}}(\mathcal{P}^{(t)}|\boldsymbol{H}^{(t)}_\mathsf{est})$ to approximate the original distribution $p(\mathcal{P}^{(t)}|\boldsymbol{H}^{(t)}_\mathsf{est})$. As a consequence, the training objective is to maximize the log-likelihood $\log p_{\boldsymbol{\theta}}(\mathcal{P}^{(t)}|\boldsymbol{H}^{(t)}_\mathsf{est})$. Nevertheless, direct maximization of $\log p_{\boldsymbol{\theta}}(\mathcal{P}^{(t)}|\boldsymbol{H}^{(t)}_\mathsf{est})$ is generally intractable due to latent yet inevitable physical factors, such as limited bandwidth, finite antenna aperture size, and measurement noise. These may result in the weak and ambiguous representations of $\boldsymbol{H}^{(t)}_\mathsf{est}$ for UAV EPs. We therefore introduce a latent variable $\boldsymbol{z}$ to explicitly encode intrinsic but partially unobservable features, together with a learnable distribution $q_{\boldsymbol{\phi}}(\boldsymbol{z}^{(t)}| \mathcal{P}^{(t)}, \boldsymbol{H}^{(t)}_{\mathsf{est}})$ that provides a stochastic estimate of $\boldsymbol{z}^{(t)}$. Accordingly, we rewrite $\log p_{\boldsymbol{\theta}}(\mathcal{P}^{(t)}|\boldsymbol{H}^{(t)}_\mathsf{est})$ as
\begin{align}
    \label{log p theta}
    &\log p_{\boldsymbol{\theta}}(\mathcal{P}^{(t)}|\boldsymbol{H}^{(t)}_\mathsf{est}) = \log\int p_{\boldsymbol{\theta}}(\mathcal{P}^{(t)},\boldsymbol{z}^{(t)}|\boldsymbol{H}^{(t)}_\mathsf{est})\text{d} \boldsymbol{z}^{(t)}\nonumber\\
    & = \log\int q_{\boldsymbol{\phi}}(\boldsymbol{z}^{(t)}|\mathcal{P}^{(t)},\boldsymbol{H}^{(t)}_\mathsf{est})\frac{p_{\boldsymbol{\theta}}(\mathcal{P}^{(t)},\boldsymbol{z}^{(t)}|\boldsymbol{H}^{(t)}_\mathsf{est})}{q_{\boldsymbol{\phi}}(\boldsymbol{z}^{(t)}|\mathcal{P}^{(t)},\boldsymbol{H}^{(t)}_\mathsf{est})}\text{d} \boldsymbol{z}^{(t)}\nonumber\\
    & = \log \left(\mathbb{E}_{q_{\boldsymbol{\phi}}(\boldsymbol{z}^{(t)}|\mathcal{P}^{(t)},\boldsymbol{H}^{(t)}_\mathsf{est})}\left[\frac{p_{\boldsymbol{\theta}}(\mathcal{P}^{(t)},\boldsymbol{z}^{(t)}|\boldsymbol{H}^{(t)}_\mathsf{est})}{q_{\boldsymbol{\phi}}(\boldsymbol{z}^{(t)}|\mathcal{P}^{(t)},\boldsymbol{H}^{(t)}_\mathsf{est})}\right]\right).
\end{align}
For notational brevity, we denote $\mathbb{E}_{q_{\boldsymbol{\phi}}(\boldsymbol{z}^{(t)}|\mathcal{P}^{(t)},\boldsymbol{H}^{(t)}_\mathsf{est})}(\cdot)$ by $\mathbb{E}_{q_{\boldsymbol{\phi}}}(\cdot)$ in the following analysis. Based on Jensen’s inequality, we can obtain an evidence lower bound (ELBO) of $\log p_{\boldsymbol{\theta}}(\mathcal{P}^{(t)}|\boldsymbol{H}^{(t)}_\mathsf{est})$ as
\begin{align}
    \label{ELBO}
    &\log p_{\boldsymbol{\theta}}(\mathcal{P}^{(t)}|\boldsymbol{H}^{(t)}_\mathsf{est}) \ge \mathbb{E}_{q_{\boldsymbol{\phi}}} \log \frac{p_{\boldsymbol{\theta}}(\mathcal{P}^{(t)},\boldsymbol{z}^{(t)}|\boldsymbol{H}^{(t)}_\mathsf{est})}{q_{\boldsymbol{\phi}}(\boldsymbol{z}^{(t)}|\mathcal{P}^{(t)},\boldsymbol{H}^{(t)}_\mathsf{est})}\nonumber\\
    & = \mathbb{E}_{q_{\boldsymbol{\phi}}}\left[\log p_{\boldsymbol{\theta}}(\mathcal{P}^{(t)}|\boldsymbol{z}^{(t)},\boldsymbol{H}^{(t)}_\mathsf{est}) - \log \frac{q_{\boldsymbol{\phi}}(\boldsymbol{z}^{(t)}|\mathcal{P}^{(t)},\boldsymbol{H}^{(t)}_\mathsf{est})}{p_{\boldsymbol{\theta}}(\boldsymbol{z}^{(t)}|\boldsymbol{H}^{(t)}_\mathsf{est})} \right]\nonumber\\
    & \triangleq \text{ELBO}(\mathcal{P}^{(t)},\boldsymbol{H}^{(t)}_\mathsf{est}).
\end{align}
Therefore, maximization of $\log p_{\boldsymbol{\theta}}(\mathcal{P}^{(t)}|\boldsymbol{H}^{(t)}_\mathsf{est})$ is recast as minimizing the following loss function:
\begin{align}
    \label{ELBO loss}
    \mathcal{L}_{\text{ELBO}}&(\mathcal{P}^{(t)},\boldsymbol{H}^{(t)}_\mathsf{est}) = \mathbb{E}_{q_{\boldsymbol{\phi}}}[-\log p_{\boldsymbol{\theta}}(\mathcal{P}^{(t)}|\boldsymbol{z}^{(t)},\boldsymbol{H}^{(t)}_\mathsf{est})]\nonumber\\
    &+D_{\text{KL}}(q_{\boldsymbol{\phi}}(\boldsymbol{z}^{(t)}|\mathcal{P}^{(t)},\boldsymbol{H}^{(t)}_\mathsf{est})\|p_{\boldsymbol{\theta}}(\boldsymbol{z}^{(t)}|\boldsymbol{H}^{(t)}_\mathsf{est})).
\end{align}

Note that \eqref{ELBO loss} serves as a standard training objective within the conditional VAE framework\cite{sohn2015learning}. 
While it is feasible in general, this objective includes a KL term conditioned on the ground-truth $\mathcal{P}^{(t)}$, which is not available at the inference stage. To this end, we adopt the following inference-consistent relaxations. First, we remove the dependence on 
$\mathcal{P}^{(t)}$ in the posterior, and allow the encoder to predict a posterior that depends solely on the measurements, i.e., $q_{\boldsymbol{\phi}}(\boldsymbol{z}^{(t)}|\boldsymbol{H}^{(t)}_\mathsf{est})$. This relaxation forces the encoder to extract all task-relevant information from $\boldsymbol{H}^{(t)}_\mathsf{est}$. Then, we replace the conditional prior with a global learnable prior $p_{\boldsymbol{\psi}}(\boldsymbol{z}^{(t)})$. This prior regularizes the latent space independent of 
$\boldsymbol{H}^{(t)}_\mathsf{est}$ and can approximate the aggregated posterior (e.g., via a normalizing flow). In addition, to ensure that information flows through the latent bottleneck and reconstruction relies solely on features extracted from the measurements, we condition the decoder primarily on $\boldsymbol{z}^{(t)}$, i.e., replacing $p_{\boldsymbol{\theta}}(\mathcal{P}^{(t)}|\boldsymbol{z}^{(t)},\boldsymbol{H}^{(t)}_\mathsf{est})$ with $p_{\boldsymbol{\theta}}(\mathcal{P}^{(t)}|\boldsymbol{z}^{(t)})$. Based on these considerations, the loss function \eqref{ELBO loss} can be reduced to a more concise form as:
\begin{align}
    \label{ELBO loss relax}
    \tilde{\mathcal{L}}_{\text{ELBO}}&(\mathcal{P}^{(t)},\boldsymbol{H}^{(t)}_\mathsf{est}) = \mathbb{E}_{q_{\boldsymbol{\phi}}(\boldsymbol{z}^{(t)}|\boldsymbol{H}^{(t)}_\mathsf{est})}[-\log p_{\boldsymbol{\theta}}(\mathcal{P}^{(t)}|\boldsymbol{z}^{(t)})]\nonumber\\
    &+D_{\text{KL}}(q_{\boldsymbol{\phi}}(\boldsymbol{z}^{(t)}|\boldsymbol{H}^{(t)}_\mathsf{est})\|p_{\boldsymbol{\psi}}(\boldsymbol{z}^{(t)})).
\end{align}
The resulting inference-aligned objective removes training–inference shortcuts, and
matches the expected inference pathway $\boldsymbol{H}^{(t)}_\mathsf{est}\to\boldsymbol{z}^{(t)}\to \mathcal{P}^{(t)}$. From \eqref{ELBO loss relax}, the proposed AUGUST approach can be decomposed into two modules:
\begin{itemize}
    \item \textbf{Conditional Channel Encoding Module}: The channel encoder $q_{\boldsymbol{\phi}}(\boldsymbol{z}^{(t)}|\boldsymbol{H}^{(t)}_\mathsf{est})$ explicitly extracts intrinsic features of the UAV into the latent code $\boldsymbol{z}^{(t)}$ only from the measurement data, thereby ensuring consistency between training and inference. Following the principle of VAE, the conditional information used for the reconstruction is sampled from a reparameterized latent distribution of $\boldsymbol{z}^{(t)}$ instead of a deterministic feature code. Moreover, a learnable global prior $p_{\boldsymbol{\psi}}(\boldsymbol{z}^{(t)})$ that is independent of the measurement is utilized to regularize the latent space through the KL divergence term.
    \item \textbf{Generative Decoding (Reconstruction) Module}: Given the latent code $\boldsymbol{z}^{(t)}$ capturing the high-dimensional features from the channel, the generative decoder $p_{\boldsymbol{\theta}}(\mathcal{P}^{(t)}|\boldsymbol{z}^{(t)})$ employs the diffusion model to reconstruct the 5D point cloud representation of the UAV, thereby achieving the unified sensing and tracking its position, attitude, and shape.
\end{itemize}
{\color{blue} It is worth noting that the learnable prior $p_{\boldsymbol{\psi}}(\boldsymbol{z}^{(t)})$ is not used as an additional conditioning input during inference. Its role is to shape the latent space during training through the KL divergence term
$D_{\text{KL}}(q_{\boldsymbol{\phi}}(\boldsymbol{z}^{(t)}|\boldsymbol{H}^{(t)}_\mathsf{est})
\|p_{\boldsymbol{\psi}}(\boldsymbol{z}^{(t)})).$ Although this prior is not an explicit inference-time input, it affects inference indirectly because the encoder used during inference is trained under this regularized latent geometry. Compared with directly using deterministic channel features without latent-space regularization, this stochastic latent bottleneck constrains the channel-induced latent codes to lie on a shared and regularized latent manifold, which helps suppress noise-induced perturbations and redundant channel variations. Moreover, compared with a fixed standard Gaussian prior, the learnable flow-based prior starts from a Gaussian base distribution but transforms it through invertible mappings to better approximate the aggregated posterior induced by diverse UAV positions, attitudes, shapes, EP parameters, and channel conditions. This reduces the posterior--prior mismatch and avoids over-constraining the task-relevant latent representation, thereby providing more consistent conditional information for the subsequent diffusion-based reconstruction module\cite{rezende2015variational}.}

\begin{figure*}[htb]
    \centering
    \begin{subfigure}[b]{.97\linewidth}
        \centering
        \includegraphics[height=8.2cm]{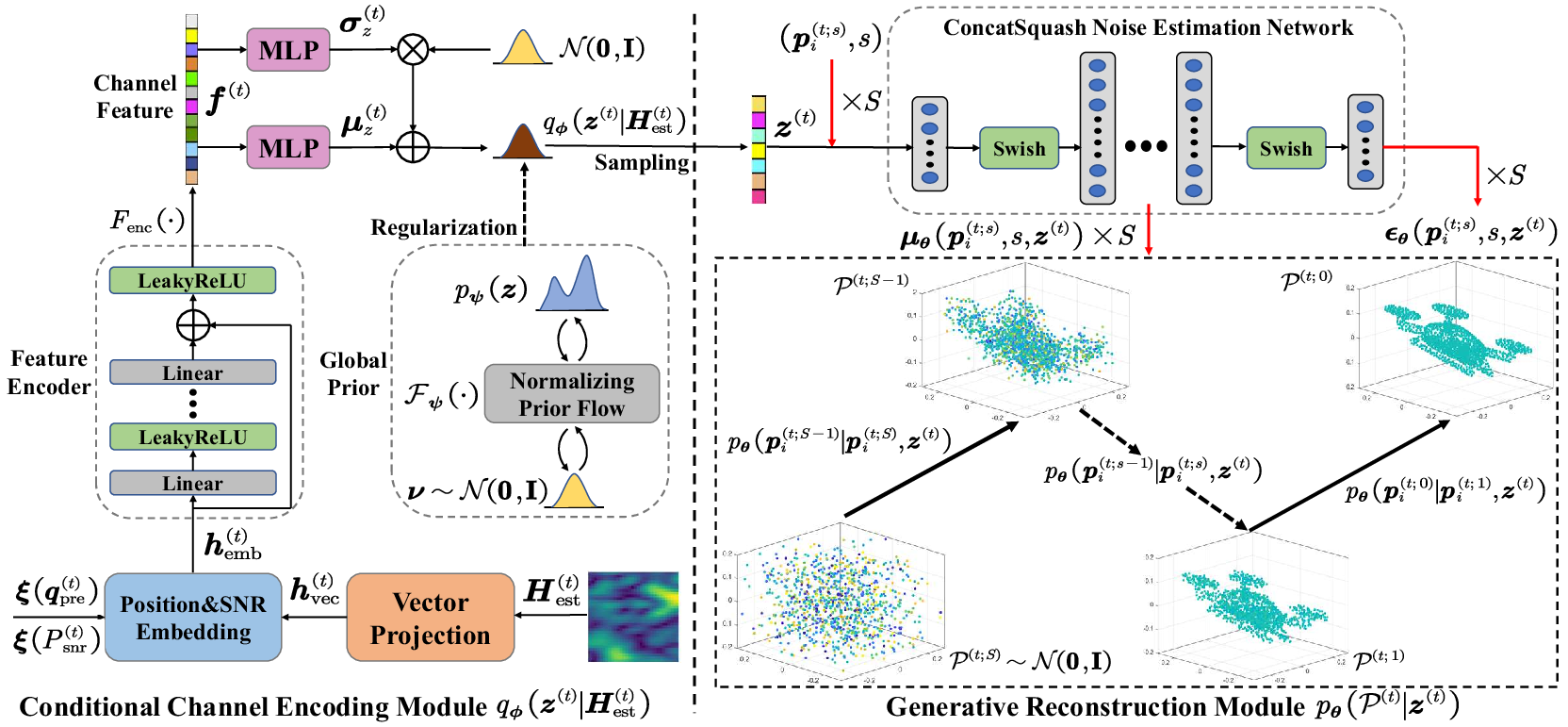}
    \end{subfigure}
    \caption{The illustration of the proposed AUGUST approach within slot $t$.}
    \label{AUGUST}
\end{figure*}

The overall description of the proposed AUGUST approach is illustrated in Fig. \ref{AUGUST}. In the subsequent subsections, we will introduce the conditional channel encoding module and the diffusion-based reconstruction module in detail.
\subsection{Conditional Channel Encoding Module}
As input to reconstruction, the sensing channel varies not only with changes in the UAV flight status but also depends strongly on the UAV’s spatial location and transmission strategies (e.g., beamforming design and transmit power). In particular, the phase associated with each scattering point follows an exponential phase mapping with respect to its spatial coordinates. Consequently, the sensing channel $\boldsymbol{H}^{(t)}_{\mathsf{est}}$ can vary rapidly with position changes due to multipath-induced phase variations. Thus, to facilitate stable UAV feature extraction by the encoder under continuously varying UAV positions and channel conditions, we introduce the positional embedding and SNR embedding before the channel encoder.

Specifically, we first vectorize the obtained $\boldsymbol{H}^{(t)}_\mathsf{est}$ and project it linearly into a $d_p$-dimensional channel vector based on a fully connected layer with parameters $\boldsymbol{W}_p$ and $\boldsymbol{b}_p$, i.e.,
\begin{align}
    \label{h vec}
    \boldsymbol{h}^{(t)}_\text{vec} = \boldsymbol{W}_p\operatorname{vec}(\boldsymbol{H}^{(t)}_\mathsf{est}) +\boldsymbol{b}_p.
\end{align}
In the considered UAV tracking scenario, the UAV point cloud $\mathcal{P}^{(t)}$ is reconstructed within a RoI centered at the predicted position for each slot $t$. Thus, the position encoding function $\boldsymbol{\xi}(\boldsymbol{q}_{\mathsf{pre}}^{(t)})$ and the SNR encoding function $\boldsymbol{\xi}(P^{(t)}_{\text{snr}})$ are employed to address the issue of capturing high-frequency variations in the sensing channel. $\boldsymbol{\xi}(\boldsymbol{q}_{\mathsf{pre}}^{(t)})$ and $\boldsymbol{\xi}(P^{(t)}_{\text{snr}})$ are defined as mappings from the predicted coordinate of the UAV and the SNR at the BS into higher-dimensional Fourier feature spaces, i.e.,
$\mathbb{R}^{3(2d_\xi+1)}$ for the position encoding and
$\mathbb{R}^{2d_\xi+1}$ for the SNR encoding, respectively, i.e.,
\begin{align}
    \label{position encoding}
    \boldsymbol{\xi}(\boldsymbol{q}_{\mathsf{pre}}^{(t)}) = &[\boldsymbol{q}_{\mathsf{pre}}^{(t)};\sin(2^0\pi \boldsymbol{q}_{\mathsf{pre}}^{(t)});\cos(2^0\pi \boldsymbol{q}_{\mathsf{pre}}^{(t)});\cdots;\nonumber\\
    &\sin(2^{d_\xi-1}\pi \boldsymbol{q}_{\mathsf{pre}}^{(t)});\cos(2^{d_\xi-1}\pi \boldsymbol{q}_{\mathsf{pre}}^{(t)})],
\end{align}
\begin{align}
    \label{SNR encoding}
    \boldsymbol{\xi}(P^{(t)}_{\text{snr}}) = &[P^{(t)}_{\text{snr}},\sin(2^0\pi P^{(t)}_{\text{snr}}),\cos(2^0\pi P^{(t)}_{\text{snr}}),\cdots,\nonumber\\
    &\sin(2^{d_\xi-1}\pi P^{(t)}_{\text{snr}}),\cos(2^{d_\xi-1}\pi P^{(t)}_{\text{snr}})],
\end{align}
where $d_\xi$ is the number of encoding frequencies, and the SNR is given by $P^{(t)}_{\text{snr}} = \frac{\|\boldsymbol{H}^{(t)}\boldsymbol{X}^{(t)}\|^2_F}{\|\boldsymbol{N}\|^2_F}$. After this transform, high-frequency spatial and value variations become two linear combinations of fixed harmonic bases. Besides, since wireless channels are continuous physical fields whose statistics are strongly coupled to the UAV position, the conventional additive positional embeddings applied in natural language processing \cite{ke2020rethinking} are therefore less suitable for the considered problem. Motivated by \cite{xing2025multi}, we then adopt a multiplicative strategy for
positional embedding, and the embedded channel vector input to the encoder is given by
\begin{align}
    \label{h emb}
    \boldsymbol{h}^{(t)}_\text{emb} = [\boldsymbol{\omega}_1(\boldsymbol{\xi}(\boldsymbol{q}_{\mathsf{pre}}^{(t)}))\odot \boldsymbol{h}^{(t)}_\text{vec};\boldsymbol{\omega}_2(\boldsymbol{\xi}(P^{(t)}_{\text{snr}}))],
\end{align}
where $\boldsymbol{\omega}_1(\cdot)$ and $\boldsymbol{\omega}_2(\cdot)$ are two shallow fully connected heads that explicitly capture the position- and SNR-dependent effects, while $\boldsymbol{h}^{(t)}_\text{vec}$ focuses on target-related information that is comparatively stable across varying UAV positions and channel conditions.

As shown in the left part of Fig. \ref{AUGUST}, given $\boldsymbol{h}^{(t)}_\text{emb}$, we next extract a feature vector $\boldsymbol{f}^{(t)}$ with dimension $d_f$ from this embedded channel vector via a feature encoder $F_{\text{enc}}$, i.e., $\boldsymbol{f}^{(t)} = F_{\text{enc}}(\boldsymbol{h}^{(t)}_\text{emb})$. Here, $F_{\text{enc}}(\cdot)$ is implemented based on a residual MLP with Linear–LeakyReLU pairs. Notably, the extracted feature vector $\boldsymbol{f}^{(t)}$ encodes the intrinsic properties of the UAV, but it is high-dimensional and deterministic, which may still contain redundant information and noise. To achieve a high-quality reconstruction and improve the generalization of the model, we further compress the dimension of $\boldsymbol{f}^{(t)}$ by using two MLPs to map $\boldsymbol{f}^{(t)}$ to two $d_z$-dimensional parameters of a Gaussian posterior: $\boldsymbol{\mu}^{(t)}_z$ and $\boldsymbol{\sigma}^{(t)}_z$. In this way, the posterior distribution of the aforementioned latent code $\boldsymbol{z}^{(t)}$ is given by
\begin{align}
    \label{q alpha z}
    q_{\boldsymbol{\phi}}(\boldsymbol{z}^{(t)}|\boldsymbol{H}^{(t)}_\mathsf{est}) = \mathcal{N}\left(\boldsymbol{z}^{(t)};\boldsymbol{\mu}^{(t)}_z,\text{diag}[\boldsymbol{\sigma}^{(t)}_z]\right).
\end{align}
Given the distribution $q_{\boldsymbol{\phi}}(\boldsymbol{z}^{(t)}|\boldsymbol{H}^{(t)}_\mathsf{est})$, the latent code $\boldsymbol{z}^{(t)}$, which is used as a conditional input for the following generative reconstruction module, can be sampled from $q_{\boldsymbol{\phi}}(\boldsymbol{z}^{(t)}|\boldsymbol{H}^{(t)}_\mathsf{est})$.

\subsection{Diffusion-Based Reconstruction Module}
Based on the obtained $\boldsymbol{z}^{(t)}$ that encodes the high-dimensional features from the sensing channel, we aim to design the decoder $p_{\boldsymbol{\theta}}(\mathcal{P}^{(t)}|\boldsymbol{z}^{(t)})$ by employing the diffusion model \cite{luo2021diffusion}, as detailed in the following. 

\textit{1) Forward Diffusion Process}: To clearly present the diffusion and denoising processes, $\mathcal{P}^{(t;1)}, \ldots, \mathcal{P}^{(t;S)}$ are denoted as a sequence of point clouds sequentially generated from the original point cloud $\mathcal{P}^{(t;0)} = \{\boldsymbol{p}^{(t;0)}_i\}^{M}_{i=1}$ defined in \eqref{5D pc}, up to the maximum time step $S$; each step is indexed by $s$. Since the 5D point cloud $\mathcal{P}^{(t;0)} = \{\boldsymbol{p}^{(t;0)}_i\}^{M}_{i=1}$ reflects the holistic distribution of UAV EPs, $p_{\boldsymbol{\theta}}(\mathcal{P}^{(t;0)}|\boldsymbol{z}^{(t)})$ can be factorized into $M$ independent samples drawn from the point distribution $q(\boldsymbol{p}^{(t;0)}_i|\boldsymbol{z}^{(t)})$. In the forward diffusion process, the transition from the original distribution $q(\boldsymbol{p}^{(t;0)}_i)$ to a noise-like one $q(\boldsymbol{p}^{(t;S)}_i)$ can be modeled as a Markov chain:
\begin{align}
    \label{forward markov chain}
 q(\boldsymbol{p}^{(t;1:S)}_i&|\boldsymbol{p}^{(t;0)}_i)  = {\prod\nolimits_{s=1}^{S}}  q(\boldsymbol{p}^{(t;s)}_i|\boldsymbol{p}^{(t;s-1)}_i)\nonumber\\
    &={\prod\nolimits_{s=1}^{S}}\mathcal{N}(\boldsymbol{p}^{(t;s)}_i;\sqrt{1-\beta_s}\boldsymbol{p}^{(t;s-1)}_i,\beta_s\mathbf{I}).
\end{align}
Here, $\beta_s$ is a hyperparameter controlling the noise intensity and increases linearly with the step index $s$. Then, according to Bayes’ rule and using the renormalization technique, the calculation of $\boldsymbol{p}^{(t;s)}_i$ at an
arbitrary time step $s$ can be simplified to
\begin{align}
    \label{forward q s}
    q(\boldsymbol{p}^{(t;s)}_i|\boldsymbol{p}^{(t;0)}_i)  = \mathcal{N}(\boldsymbol{p}^{(t;s)}_i;\sqrt{\overline{\alpha}_s}\boldsymbol{p}^{(t;0)}_i,\sqrt{1-\overline{\alpha}_s}\mathbf{I}),
\end{align}
where $\alpha_s = 1-\beta_s$ and $\overline{\alpha}_s = \prod^{s}_{s'=1}\alpha_{s'} = {\alpha}_s\overline{\alpha}_{s-1}$.

\textit{2) Reverse Diffusion Process}: Under the diffusion framework, the reconstruction process corresponds to the reverse of the forward diffusion process. From \eqref{forward markov chain} and \eqref{forward q s}, the reverse transition probability conditioned on $\boldsymbol{p}^{(t;0)}_i$ is given by
\begin{align}
    \label{reverse q s}
    q(\boldsymbol{p}^{(t;s-1)}_i|\boldsymbol{p}^{(t;s)}_i,\boldsymbol{p}^{(t;0)}_i)  = \mathcal{N}(\boldsymbol{p}^{(t;s-1)}_i;\tilde{\boldsymbol{\mu}}_s (\boldsymbol{p}^{(t;s)}_i,\boldsymbol{p}^{(t;0)}_i),\tilde{\beta}_s\mathbf{I}),
\end{align}
with 
\begin{align}
    \label{mu s}
\tilde{\boldsymbol{\mu}}_s (\boldsymbol{p}^{(t;s)}_i,\boldsymbol{p}^{(t;0)}_i) & = \frac{\sqrt{\overline{\alpha}_{s-1}}\beta_s}{1-\overline{\alpha}_{s}}\boldsymbol{p}^{(t;0)}_i+\frac{\sqrt{\alpha_s}(1-\overline{\alpha}_{s-1})}{1-\overline{\alpha}_{s}}\boldsymbol{p}^{(t;s)}_i\nonumber\\
& = \frac{1}{\sqrt{\alpha_s}}\left(\boldsymbol{p}^{(t;s)}_i-\tilde{\alpha}_s\boldsymbol{\epsilon}_i^{(s)}\right).
\end{align}
Here, $\tilde{\beta}_s = \frac{1-\overline{\alpha}_{s-1}}{1-\overline{\alpha}_{s}}\beta_s$, $\tilde{\alpha}_s=\frac{1-\alpha_s}{\sqrt{1-\overline{\alpha}_{s}}}$, and $\boldsymbol{\epsilon}_i^{(s)} = \frac{1}{\sqrt{1-\overline{\alpha}_{s}}}(\boldsymbol{p}^{(t;s)}_i-\sqrt{\overline{\alpha}_{s}}\boldsymbol{p}^{(t;0)}_i)$. However, during reconstruction, the original point cloud $\mathcal{P}^{(t;0)}$ is unknown, and the only available conditioning information is $\boldsymbol{z}^{(t)}$ encoded from the sensing channel. To tackle this, the reverse process is implemented by introducing a sequence of learnable, parameterized transition posteriors conditioned on $\boldsymbol{z}^{(t)}$, i.e.,
\begin{align}
    \label{reverse markov chain}
    p_{\boldsymbol{\theta}}(\boldsymbol{p}^{(t;0:S)}_i|\boldsymbol{z}^{(t)}) = p(\boldsymbol{p}^{(t;S)}_i)\prod\nolimits_{s=1}^{S} p_{\boldsymbol{\theta}}(\boldsymbol{p}^{(t;s-1)}_i|\boldsymbol{p}^{(t;s)}_i,\boldsymbol{z}^{(t)}).
\end{align}
Here, $p(\boldsymbol{p}^{(t;S)}_i)$ is a predetermined Gaussian distribution from which the initial random point cloud $\mathcal{P}^{(t;S)}$ is sampled. $p_{\boldsymbol{\theta}}(\boldsymbol{p}^{(t;s-1)}_i|\boldsymbol{p}^{(t;s)}_i,\boldsymbol{z}^{(t)})$ denotes the learnable reverse transition probability that follows the Gaussian distribution:
\begin{align}
    \label{reverse p theta}
    p_{\boldsymbol{\theta}}(\boldsymbol{p}^{(t;s-1)}_i|\boldsymbol{p}^{(t;s)}_i,\boldsymbol{z}^{(t)}) = \mathcal{N}(\boldsymbol{p}^{(t;s-1)}_i;{\boldsymbol{\mu}}_{\boldsymbol{\theta}} (\boldsymbol{p}^{(t;s)}_i,s,\boldsymbol{z}^{(t)}),\tilde{\beta}_s\mathbf{I}).
\end{align}
Since \eqref{reverse p theta} is a parameterized approximation of \eqref{reverse q s}, ${\boldsymbol{\mu}}_{\boldsymbol{\theta}} (\boldsymbol{p}^{(t;s)}_i,s,\boldsymbol{z}^{(t)})$ has a similar form as $\tilde{\boldsymbol{\mu}}_s (\boldsymbol{p}^{(t;s)}_i,\boldsymbol{p}^{(t;0)}_i)$, which is given by
\begin{align}
    \label{mu theta}
{\boldsymbol{\mu}}_{\boldsymbol{\theta}} (\boldsymbol{p}^{(t;s)}_i,s,\boldsymbol{z}^{(t)}) = \frac{1}{\sqrt{\alpha_s}}\left(\boldsymbol{p}^{(t;s)}_i-\tilde{\alpha}_s\boldsymbol{\epsilon}_{\boldsymbol{\theta}}(\boldsymbol{p}^{(t;s)}_i,s,\boldsymbol{z}^{(t)})\right),
\end{align}
where $\boldsymbol{\epsilon}_{\boldsymbol{\theta}}(\boldsymbol{p}^{(t;s)}_i,s,\boldsymbol{z}^{(t)})$ denotes the learning-dependent noise estimator. To implement this noise predictor, we adopt a unified noise estimation network composed of a series of $L_{\text{cs}}$ ConcatSquash layers proposed in \cite{grathwohl2018ffjord}, and each layer can be formulated as:
\begin{align}
    \label{concatsquash}
    \boldsymbol{\eta}^{\ell+1} = (\boldsymbol{W}^{\ell}_1\boldsymbol{\eta}^{\ell}+\boldsymbol{b}^{\ell}_1)\odot\varsigma(\boldsymbol{W}^{\ell}_2\boldsymbol{c}^{(t)}+\boldsymbol{b}^{\ell}_2)+\boldsymbol{W}^{\ell}_3\boldsymbol{c}^{(t)}.
\end{align}
Here, $\boldsymbol{\eta}^{\ell}$ and $\boldsymbol{\eta}^{\ell+1}$ are the input and output of the $\ell$-th layer. $\varsigma(\cdot)$ is the Sigmoid function. $\boldsymbol{W}^{\ell}_1$, $\boldsymbol{W}^{\ell}_2$, $\boldsymbol{W}^{\ell}_3$, $\boldsymbol{b}^{\ell}_1$, and $\boldsymbol{b}^{\ell}_2$ denote the trainable weights and biases in the network. The context vector $\boldsymbol{c}^{(t)} = [\overline{s},\sin(\overline{s}),\cos({\overline{s}}),\sin(2\overline{s}),\cos(2\overline{s});(\boldsymbol{z}^{(t)})^\mathsf{T}]^\mathsf{T}$ incorporates the embedded features of the normalized diffusion step $\overline{s} = \frac{s}{S}$ and the latent code $\boldsymbol{z}^{(t)}$. Swish function is used as the activation function between two adjacent layers. The input of the whole network is the 5D point $\boldsymbol{p}^{(t;s)}_i$, and the output is the estimated noise $\boldsymbol{\epsilon}_{\boldsymbol{\theta}}(\boldsymbol{p}^{(t;s)}_i,s,\boldsymbol{z}^{(t)})$.

\textit{3) ELBO of the Diffusion Model}: In the reverse diffusion process, the training goal is to maximize the log-likelihood $\log p_{\boldsymbol{\theta}}(\mathcal{P}^{(t;0)}|\boldsymbol{z}^{(t)})$ given the original UAV point cloud and the latent code $\boldsymbol{z}^{(t)}$. Clearly, direct maximization of this log-likelihood is challenging. Thus, we introduce the forward Markov chain as in \eqref{forward markov chain} to construct an ELBO of $\log p_{\boldsymbol{\theta}}(\mathcal{P}^{(t;0)}|\boldsymbol{z}^{(t)})$:
\begin{align}
    \label{ELBO log p theta}
   & \log p_{\boldsymbol{\theta}}(\mathcal{P}^{(t;0)}|\boldsymbol{z}^{(t)}) = \log \left[\mathbb{E}_{q(\mathcal{P}^{(t;1:S)}|\mathcal{P}^{(t;0)})}\frac{p_{\boldsymbol{\theta}}(\mathcal{P}^{(t;0:S)}|\boldsymbol{z}^{(t)})}{q(\mathcal{P}^{(t;1:S)}|\mathcal{P}^{(t;0)})}\right]\nonumber\\
    &\ge\mathbb{E}_{q(\mathcal{P}^{(t;1:S)}|\mathcal{P}^{(t;0)})} \log \left(\frac{p_{\boldsymbol{\theta}}(\mathcal{P}^{(t;0:S)}|\boldsymbol{z}^{(t)})}{q(\mathcal{P}^{(t;1:S)}|\mathcal{P}^{(t;0)})}\right)\nonumber\\
    & = \mathbb{E}_{q(\mathcal{P}^{(t;1:S)}|\mathcal{P}^{(t;0)})} \left[ \sum^{M}_{i=1} \log \left(\frac{p_{\boldsymbol{\theta}}(\boldsymbol{p}_i^{(t;0:S)}|\boldsymbol{z}^{(t)})}{q(\boldsymbol{p}_i^{(t;1:S)}|\boldsymbol{p}_i^{(t;0)})}\right)\right].
\end{align}
Note that \eqref{ELBO log p theta} rewrites the intractable log-likelihood $ \log p_{\boldsymbol{\theta}}(\mathcal{P}^{(t;0)}|\boldsymbol{z}^{(t)})$ as a tractable lower bound by taking an expectation over an easy-to-sample surrogate path distribution $q(\mathcal{P}^{(t;1:S)}|\mathcal{P}^{(t;0)})$. Such a formulation implicitly encourages the learned reverse process to match the true posterior over the diffusion path. In the following, we introduce a weighted training design for the proposed AUGUST approach.

\section{Weighted Training Design}
The proposed AUGUST framework is designed to reconstruct the original UAV point cloud based on the estimated channel at each time slot $t$ during tracking. Thus, the overall training objective is to maximize $\mathbb{E}_{p_{\text{train}}(\mathcal{P}^{(0:S)},\boldsymbol{H}_\mathsf{est})}[\log p_{\boldsymbol{\theta}}(\mathcal{P}^{(0)}|\boldsymbol{H}_\mathsf{est})]$. Here, $p_{\text{train}}(\mathcal{P}^{(0:S)},\boldsymbol{H}_\mathsf{est})$ denotes the distribution of the training data, including the point cloud series and the estimated channel. To this end, by substituting the ELBO in \eqref{ELBO log p theta} into the loss $\tilde{\mathcal{L}}_{\text{ELBO}}$ in \eqref{ELBO loss relax}, the training loss can be reformulated as 
\begin{align}
    \label{ELBO loss relax 2}
  \tilde{\mathcal{L}}_{\text{train}} & = \mathbb{E}_{p_{\text{train}}(\mathcal{P}^{(0:S)},\boldsymbol{H}_\mathsf{est})}\left\{ D_{\text{KL}}(q_{\boldsymbol{\phi}}(\boldsymbol{z}|\boldsymbol{H}_\mathsf{est})\|p_{\boldsymbol{\psi}}(\boldsymbol{z}))\right.\nonumber\\
   &\left.+\mathbb{E}_{\boldsymbol{z},\mathcal{P}^{(1:S)}} \left[ \sum^{M}_{i=1} \log \left(\frac{p_{\boldsymbol{\theta}}(\boldsymbol{p}_i^{(0:S)}|\boldsymbol{z})}{q(\boldsymbol{p}_i^{(1:S)}|\boldsymbol{p}_i^{(0)})}\right)\right]\right\}.
\end{align}
Then, by following Bayes' rule and according to the forward and reverse Markov chains in \eqref{forward markov chain} and \eqref{reverse markov chain}, the above loss function can be further recast as a combination of different loss functions:
\begin{align}
    \label{ELBO loss seperate}
    &\tilde{\mathcal{L}}_{\text{train}} = \mathbb{E}_{p_{\text{train}}}\bigg[\sum^{M}_{i=1}\underbrace{D_{\text{KL}}(q(\boldsymbol{p}^{(S)}_i|\boldsymbol{p}^{(0)}_i)\|p(\boldsymbol{p}^{(S)}_i) )}_{\mathcal{L}_{\mathcal{N}}}\nonumber\\
    &+\sum^{S}_{s=2} \underbrace{ D_{\text{KL}}(q(\boldsymbol{p}^{(s-1)}_i|\boldsymbol{p}^{(s)}_i,\boldsymbol{p}^{(0)}_i)\|p_{\boldsymbol{\theta}}(\boldsymbol{p}^{(s-1)}_i|\boldsymbol{p}^{(s)}_i,\boldsymbol{z}) )}_{\mathcal{L}_{\text{DM};i}^{(s)}}\nonumber\\
    &\underbrace{-\log p_{\boldsymbol{\theta}}(\boldsymbol{p}_i^{(0)}|\boldsymbol{p}_i^{(1)},\boldsymbol{z})}_{\mathcal{L}_{\text{DM};i}^{(1)}}+\underbrace{D_{\text{KL}}(q_{\boldsymbol{\phi}}(\boldsymbol{z}|\boldsymbol{H}_\mathsf{est})\|p_{\boldsymbol{\psi}}(\boldsymbol{z}))}_{\mathcal{L}_{\boldsymbol{z}}}\bigg].
\end{align}
Here, $\mathcal{L}_{\mathcal{N}}$ is a learning-free term about the Gaussian initialization of the reverse diffusion process, and $\mathcal{L}_{\text{DM};i}^{(1)}$ is a simplified term of $\mathcal{L}_{\text{DM};i}^{(s)}$ when $s=1$. $\mathcal{L}_{\text{DM};i}^{(s)}$ denotes the loss function of the diffusion model-based noise estimator $\boldsymbol{\epsilon}_{\boldsymbol{\theta}}(\boldsymbol{p}^{(s)}_i,s,\boldsymbol{z})$. From \eqref{reverse q s}-\eqref{mu theta}, $\mathcal{L}_{\text{DM};i}^{(s)}$ can be given by
\begin{align}
    \label{loss concatquash}
    \mathcal{L}_{\text{DM};i}^{(s)} = C_{\alpha_s,\beta_s}\|\boldsymbol{\epsilon}_i^{(s)}-\boldsymbol{\epsilon}_{\boldsymbol{\theta}}(\boldsymbol{p}^{(s)}_i,s,\boldsymbol{z})\|^2,
\end{align}
where $C_{\alpha_s,\beta_s}$ is a constant related to the noise intensity parameters, which can be ignored during the training \cite{jiang2024electromagnetic}. However, the numerical and spatial distribution complexities of the 3D positions and that for EPs in UAV point clouds exhibit significantly different distributions. To ensure the effectiveness of the reconstruction, we aim to train these two parts separately through a weighted $\mathcal{L}_{\text{DM};i}^{(s)}$ as
\begin{align}
    \label{loss concatquash weighted}
    \mathcal{L}_{\text{DM};i}^{(s)} = \gamma_{\text{pos}} \mathcal{L}_{\text{pos};i}^{(s)}+\gamma_{\text{EP}} \mathcal{L}_{\text{EP};i}^{(s)},
\end{align}
where $\mathcal{L}_{\text{pos};i}^{(s)} = \|\boldsymbol{\epsilon}_{\text{pos};i}^{(s)}-\boldsymbol{\epsilon}_{\text{pos};\boldsymbol{\theta}}(\boldsymbol{p}^{(s)}_i,s,\boldsymbol{z})\|^2$ and $\mathcal{L}_{\text{EP};i}^{(s)} = \|\boldsymbol{\epsilon}_{\text{EP};i}^{(s)}-\boldsymbol{\epsilon}_{\text{EP};\boldsymbol{\theta}}(\boldsymbol{p}^{(s)}_i,s,\boldsymbol{z})\|^2$ denote the diffusion loss for 3D position and EPs of the $i$-th 5D point, respectively. $\gamma_{\text{pos}}$ and $\gamma_{\text{EP}}$ are adjustable weighted coefficients for the corresponding loss. Their dimensions correspond to the definition in \eqref{5D pc}.
\begin{algorithm}[t]
\caption{Training Procedure of AUGUST Approach}
\label{algorithm train}
\begin{algorithmic}[1]
\Require Sampled point cloud series $\mathcal{P}^{(t;0:S)}$ with uniform time step $s\in\{1,\cdots,S\}$, and estimated sensing channel $\boldsymbol{H}^{(t)}_\mathsf{est}$ from $p_{\text{train}}(\mathcal{P}^{(0:S)},\boldsymbol{H}_\mathsf{est})$.
\Ensure Trained parameters: \((\boldsymbol{\phi},\boldsymbol{\psi},\boldsymbol{\theta})\).
\Repeat
\State Sample a batch from the data $\mathcal{P}^{(0:S)}$ and $\boldsymbol{H}_\mathsf{est}$.
\State Draw a latent code $\boldsymbol{z}\sim q_{\boldsymbol{\phi}}(\boldsymbol{z}|\boldsymbol{H}_\mathsf{est})$.
\For{\(i=1\) to \(M\) \textbf{in parallel}}
  \State Sample \(\boldsymbol{\epsilon}_i^{(s)} \sim\mathcal N(0,\mathbf I)\).
  \State Diffusion loss: $\mathcal{L}_{\text{DM};i}^{(s)} = \gamma_{\text{pos}} \mathcal{L}_{\text{pos};i}^{(s)}+\gamma_{\text{EP}} \mathcal{L}_{\text{EP};i}^{(s)}$.
\EndFor
\State Flow-prior regularization loss:
\Statex \qquad $\mathcal{L}_{\boldsymbol{z}} = D_{\text{KL}}\left(q_{\boldsymbol{\phi}}(\boldsymbol{z}|\boldsymbol{H}_\mathsf{est})\|\left|\text{det}\frac{\partial\mathcal{F}_{\boldsymbol{\psi}}}{\partial \boldsymbol{\nu}}\right|^{-1}p_{\mathcal{N}}(\boldsymbol{\nu})\right)$.
\State Parameters update based on gradient descent:
\Statex \qquad$(\boldsymbol{\phi},\boldsymbol{\psi},\boldsymbol{\theta})^{\text{itr}}-r_{l}\nabla_{\boldsymbol{\phi},\boldsymbol{\psi},\boldsymbol{\theta}}\mathcal{L}_{\text{train}}\to(\boldsymbol{\phi},\boldsymbol{\psi},\boldsymbol{\theta})^{\text{itr}+1}$
\Statex \qquad with $\mathcal{L}_{\text{train}} = \gamma_{\text{pos}} \mathcal{L}_{\text{pos};i}^{(s)}+\gamma_{\text{EP}} \mathcal{L}_{\text{EP};i}^{(s)}+\gamma_{\boldsymbol{z}}\mathcal{L}_{\boldsymbol{z}}$.
\Until{convergence}
\end{algorithmic}
\end{algorithm}

Regarding $\mathcal{L}_{\boldsymbol{z}}$ in \eqref{ELBO loss seperate}, it is a KL divergence term that regularizes the latent space by aligning the posterior $q_{\boldsymbol{\phi}}(\boldsymbol{z}|\boldsymbol{H}_\mathsf{est})$ with a global learnable prior $p_{\boldsymbol{\psi}}(\boldsymbol{z})$, thereby improving the consistency of the latent representation and preventing latent drift across slots. To accommodate the high-dimensional and complex distribution of $q_{\boldsymbol{\phi}}(\boldsymbol{z}|\boldsymbol{H}_\mathsf{est})$, a learnable normalizing flow $\mathcal{F}_{\boldsymbol{\psi}}$ is adopted to parameterize the prior, which reformulates $\mathcal{L}_{\boldsymbol{z}}$ without changing its regularization objective. Concretely, $\mathcal{F}_{\boldsymbol{\psi}}$ is a differentiable bijective mapping satisfying $\mathcal{F}_{\boldsymbol{\psi}}(\boldsymbol{\nu})=\boldsymbol{z}$ and $\boldsymbol{\nu}=\mathcal{F}^{-1}_{\boldsymbol{\psi}}(\boldsymbol{z})$. Then, according to \textit{Change-of-Variables Theorem}\cite{dinh2017density}, $\mathcal{L}_{\boldsymbol{z}}$ can be expressed as
\begin{align}
    \label{loss z}
    \mathcal{L}_{\boldsymbol{z}} &= D_{\text{KL}}\left(q_{\boldsymbol{\phi}}(\boldsymbol{z}|\boldsymbol{H}_\mathsf{est})\|p_{\boldsymbol{\psi}}(\boldsymbol{z})\right)\nonumber\\
    &\triangleq D_{\text{KL}}\left(q_{\boldsymbol{\phi}}(\boldsymbol{z}|\boldsymbol{H}_\mathsf{est})\|\left|\text{det}\frac{\partial\mathcal{F}_{\boldsymbol{\psi}}}{\partial \boldsymbol{\nu}}\right|^{-1}p_{\mathcal{N}}(\boldsymbol{\nu})\right).
\end{align}
Here, the prior $p_{\boldsymbol{\psi}}(\boldsymbol{z})$ is constructed by transforming a Gaussian base distribution $p_{\mathcal{N}}(\boldsymbol{\nu})=\mathcal{N}(\mathbf{0},\mathbf{I})$ through $\mathcal{F}_{\boldsymbol{\psi}}$. In this way, \eqref{loss z} enables evaluating the KL loss via the base variable $\boldsymbol{\nu}$ while retaining an expressive prior. Therefore, given \eqref{loss concatquash weighted}, \eqref{loss z}, and the uniformly sampled time step $s$, the end-to-end weighted training loss of the proposed AUGUST approach is given by:
\begin{align}
    \label{loss joint}
    \mathcal{L}_{\text{train}} = \mathbb{E}_{p_{\text{train}},s,\boldsymbol{z},\boldsymbol{\epsilon}_i^{(s)}}\{\gamma_{\text{pos}} \mathcal{L}_{\text{pos};i}^{(s)}+\gamma_{\text{EP}} \mathcal{L}_{\text{EP};i}^{(s)}+\gamma_{\boldsymbol{z}}\mathcal{L}_{\boldsymbol{z}}\},
\end{align}
where $\gamma_{\boldsymbol{z}}$ is an adjustable weighted coefficient, and the expectation is implemented based on Monte Carlo sampling and averaging for the data. The overall training procedure is summarized in \textbf{Algorithm 1}. For inference, we can obtain the reconstructed UAV point cloud (denoted as $\hat{\mathcal{P}}^{(t;0)} = \{\hat{\boldsymbol{p}}^{(t;0)}_i\}^{M}_{i=1}$) from the initial noise samples by successively sampling $p_{\boldsymbol{\theta}}(\boldsymbol{p}^{(t;0:S)}_i|\boldsymbol{z}^{(t)})$ as shown in Fig. \ref{AUGUST}.

\section{Numerical Results}
In this section, numerical experiments are conducted to evaluate the performance of the proposed approach and the considered benchmark schemes. The performance analysis of different schemes and experimental setups for UAV positioning, point cloud reconstruction quality, and attitude estimation is provided in detail.

\subsection{Experiment Setting and Benchmark Schemes}
Assume that the BS is centered at $(0,0,0)$ m, and the antenna array operates at $f_c = 3$ GHz. The noise power $\sigma^2=-120$ dBm and the transmission power is set to $P_{\mathsf{s}} \in [10,40]$ dBm. The number of transmitted symbols $L$ within each slot $t$ matches the number of antennas $N_b$ by default. To construct the UAV point cloud samples, we select five typical shapes of rotary-wing UAVs from \cite{chang2015shapenet}. We perform surface discrete random sampling and local coordinate system transformation on all UAVs to obtain UAV point cloud samples with different attitudes and shapes. The number of points in each UAV point cloud is fixed at $M=1000$ for both training and inference. In addition, a total of 50,000 UAV point cloud samples (10,000 for each shape) are generated for training, validation, and testing with a ratio of 8:1:1. Unless otherwise specified, these samples are independently and randomly generated based on 1000 random discrete trajectories and subject to the flight range $q_{\rm{x/y}}\in[10,30]$ m, $q_{\rm{z}}\in[10,20]$ m, and the tilt angle range $\theta_{\rm{x/y}}\in[-30^{\circ},30^{\circ}]$, $\theta_{\rm{z}}\in[-180^{\circ},180^{\circ}]$ as illustrated in Fig. \ref{system model}. The standard deviations of the reconstruction region are set to $s_{x/y/z}=0.85$ m. Each trajectory is composed of $T=50$ discrete points with $\Delta_t = 0.2$ s, where the UAV velocity, acceleration, and corresponding tilt angles all conform to basic constraints detailed in \cite{dai2025attitude}. Moreover, the relative permittivity and conductivity of the UAVs are set within the ranges $\varepsilon\in[1.5,5]$ and $\varrho\in[1,10]$ mS/m. The integral equations of the electromagnetic scattering process are tackled in discrete form based on methods of moments (MoM) \cite{jiang2024electromagnetic}.

\begin{figure*}[t]
        \centering
        \subcaptionbox{\footnotesize Random point cloud\label{R3 step1}}[0.196\textwidth]%
        {\includegraphics[width=\linewidth]{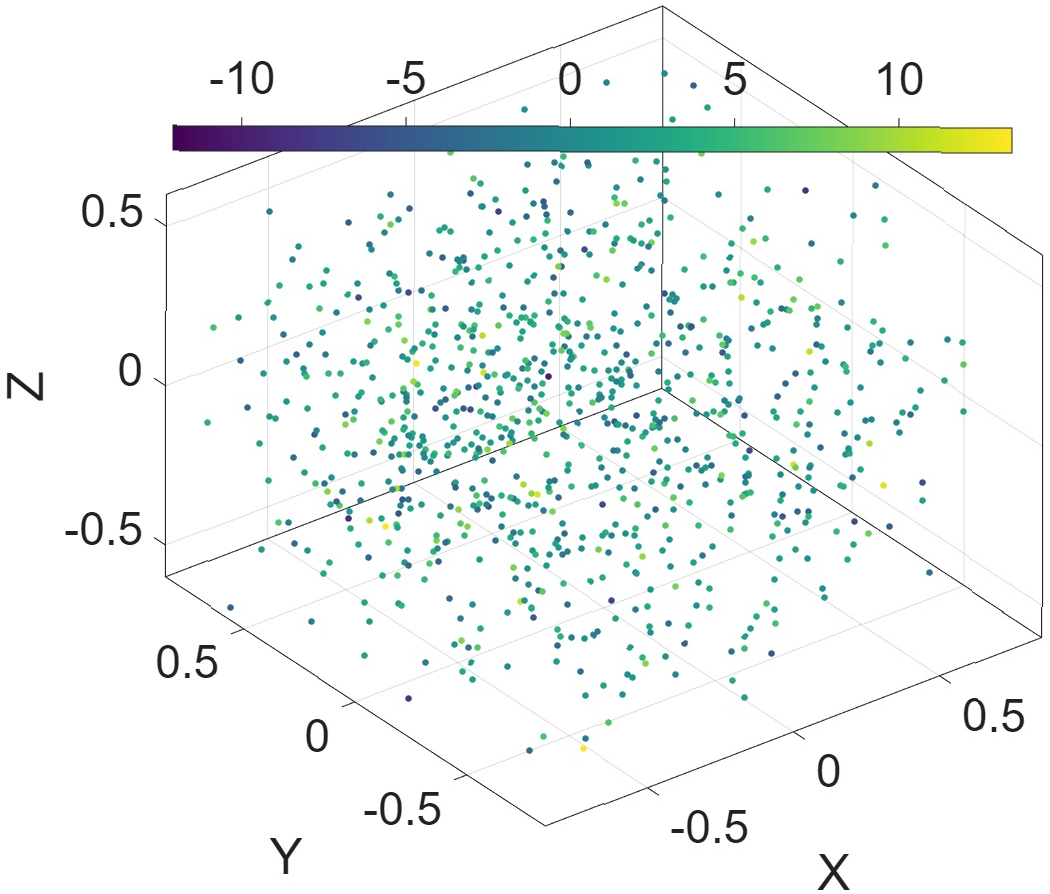}}\hfill
        \subcaptionbox{\footnotesize 80 steps\label{R3 step80}}[0.196\textwidth]%
        {\includegraphics[width=\linewidth]{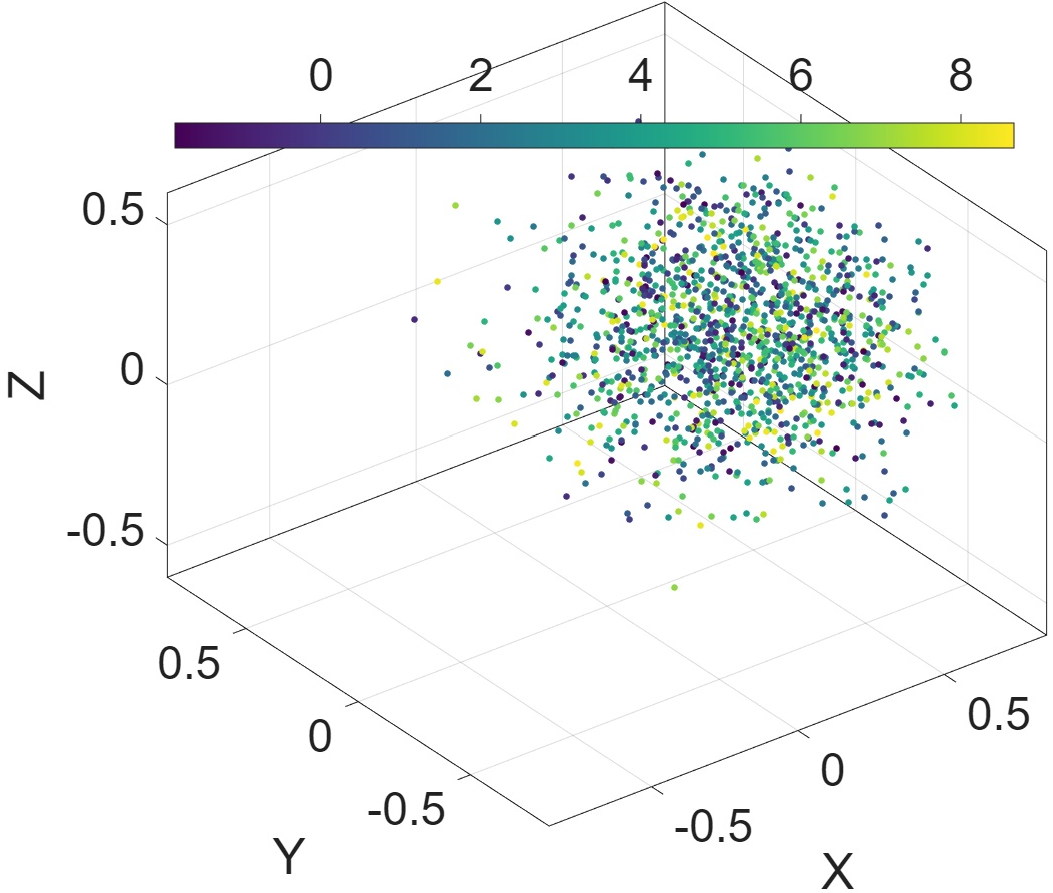}}\hfill
        \subcaptionbox{\footnotesize 150 steps\label{R3 step150}}[0.196\textwidth]%
        {\includegraphics[width=\linewidth]{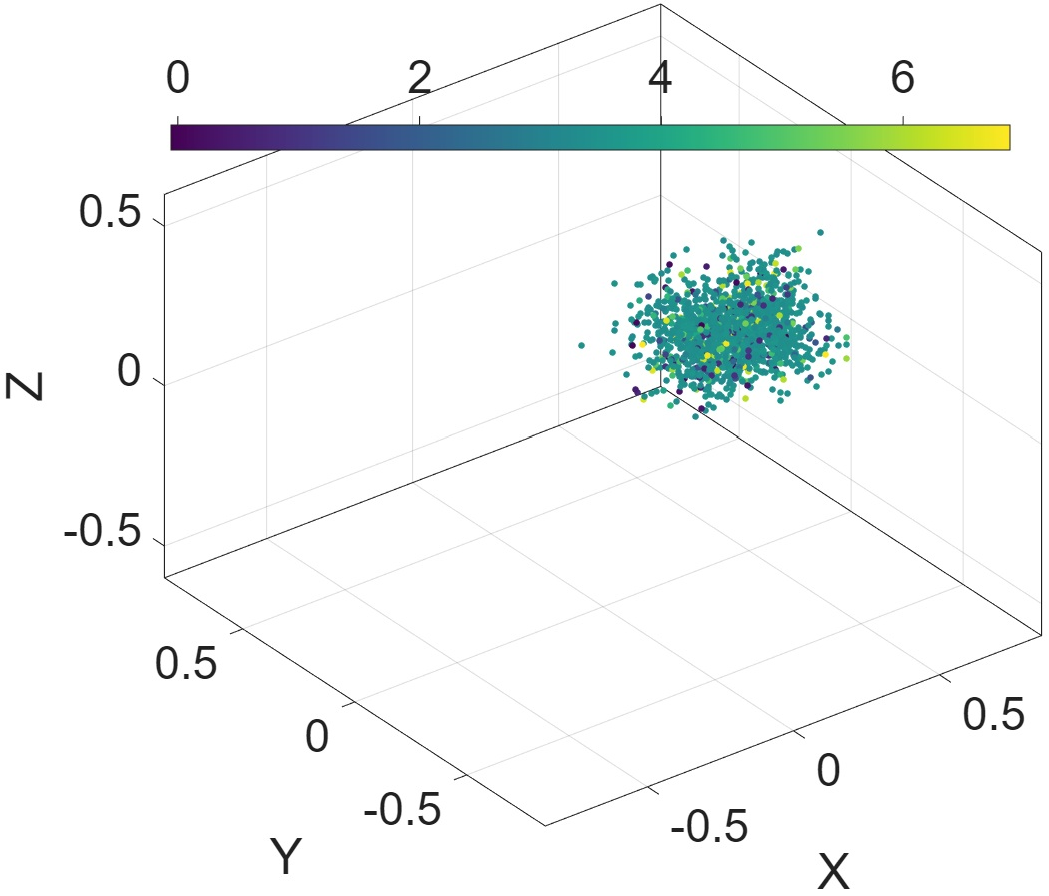}}\hfill
        \subcaptionbox{\footnotesize 180 steps\label{R3 step180}}[0.196\textwidth]%
        {\includegraphics[width=\linewidth]{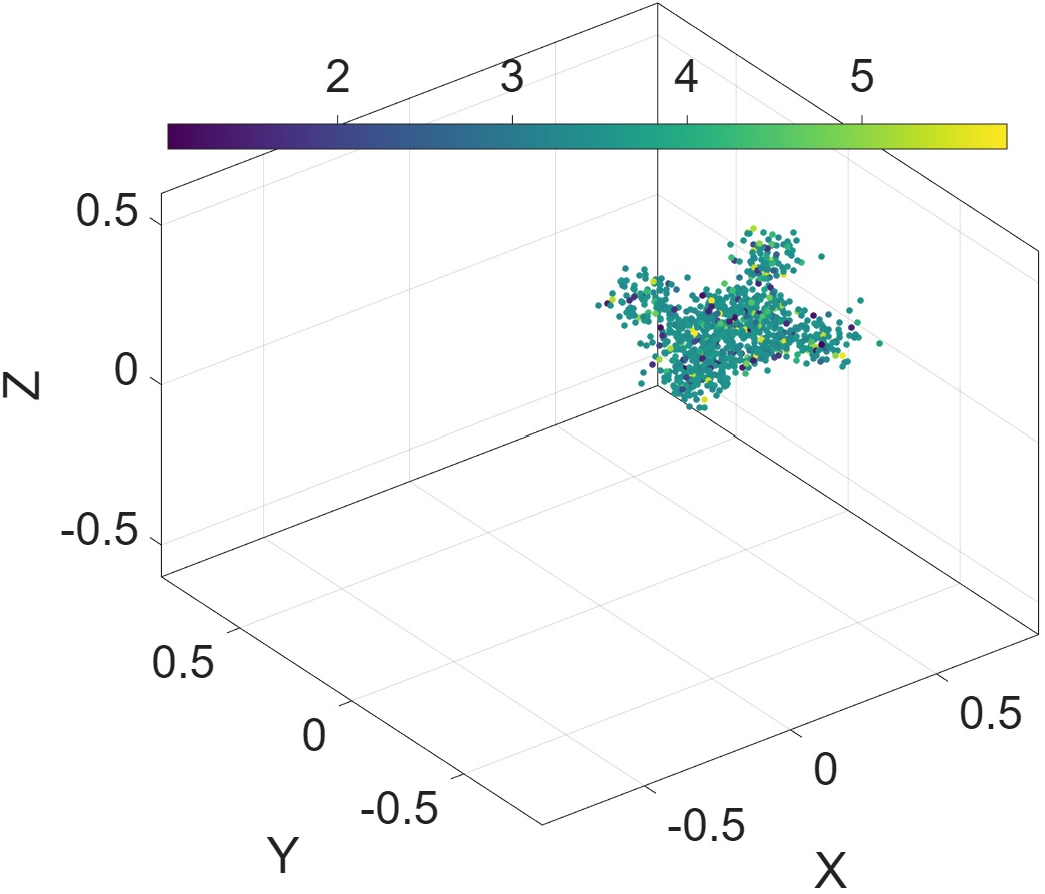}}\hfill
        \subcaptionbox{\footnotesize 200 steps\label{R3 step200}}[0.196\textwidth]%
        {\includegraphics[width=\linewidth]{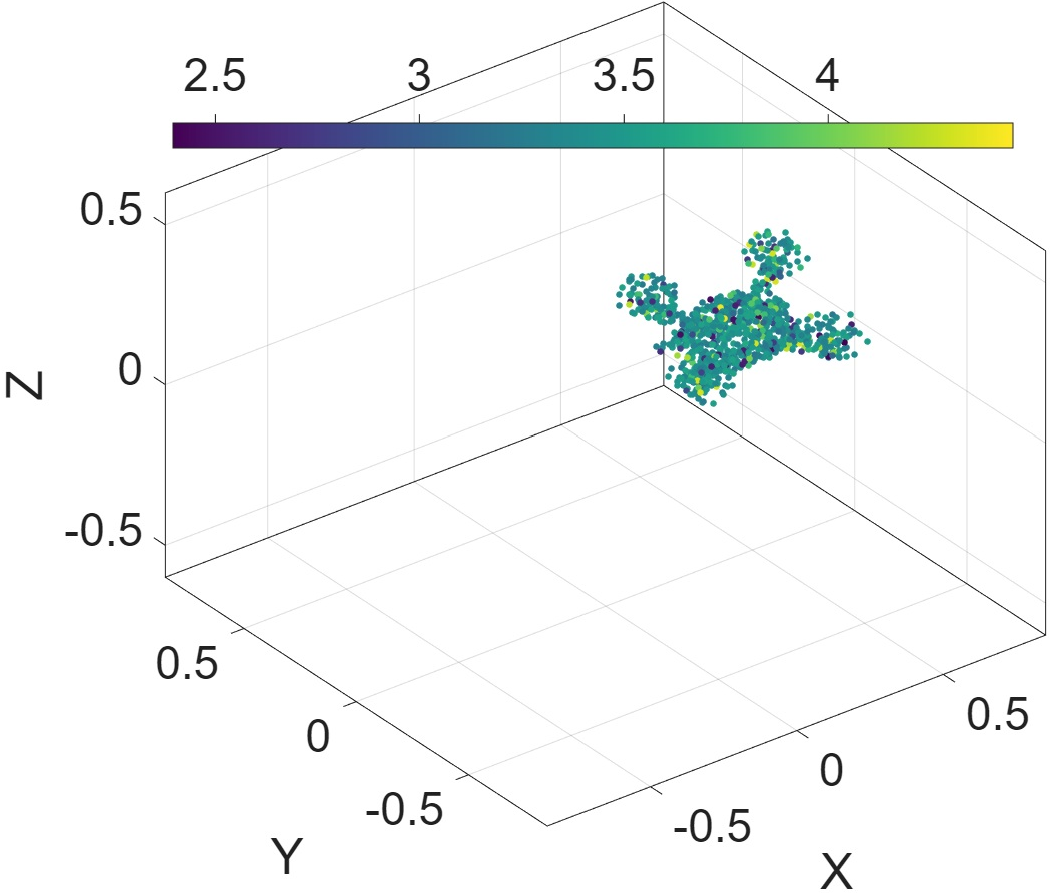}}
        \caption{\color{blue}The reverse diffusion process of EP point cloud based on the AUGUST approach, where the relative permittivity is presented.}
        \label{R3 denoise process}
    \end{figure*}

\begin{figure}[t]
\centering
\subcaptionbox{\footnotesize Ground Truth\label{PositionPre_20dB_GT}}%
{\includegraphics[width=0.24\textwidth]{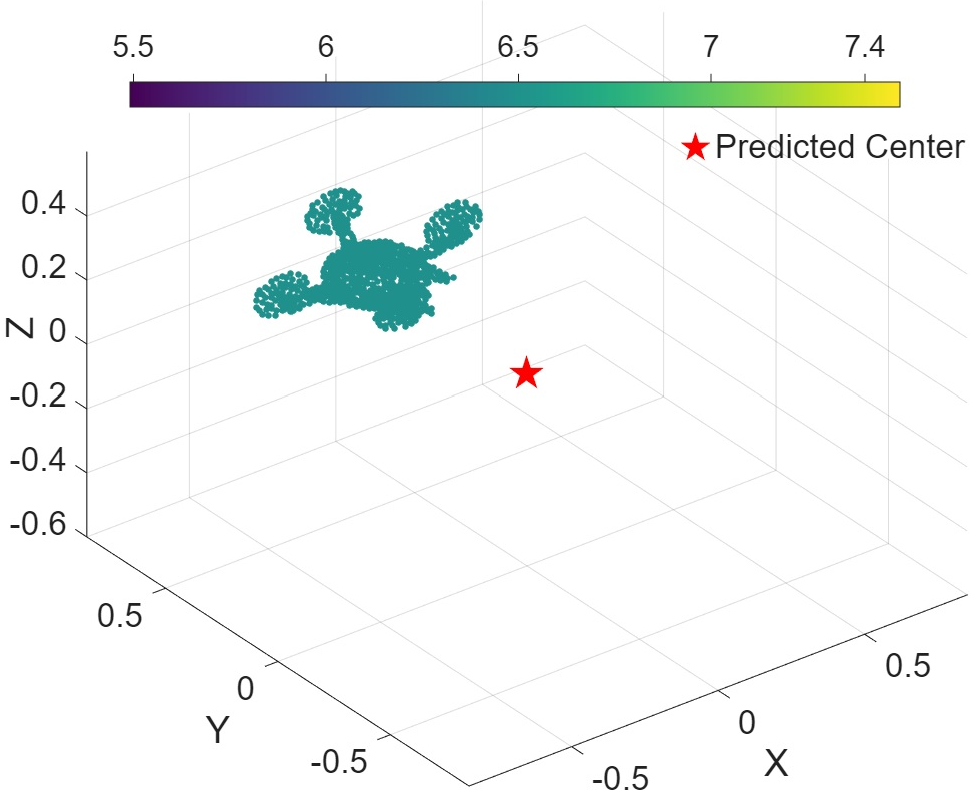}}
\subcaptionbox{\footnotesize Reconstruction ($\text{SNR}\approx 20$ dB)\label{PositionPre_20dB}}%
{\includegraphics[width=0.24\textwidth]{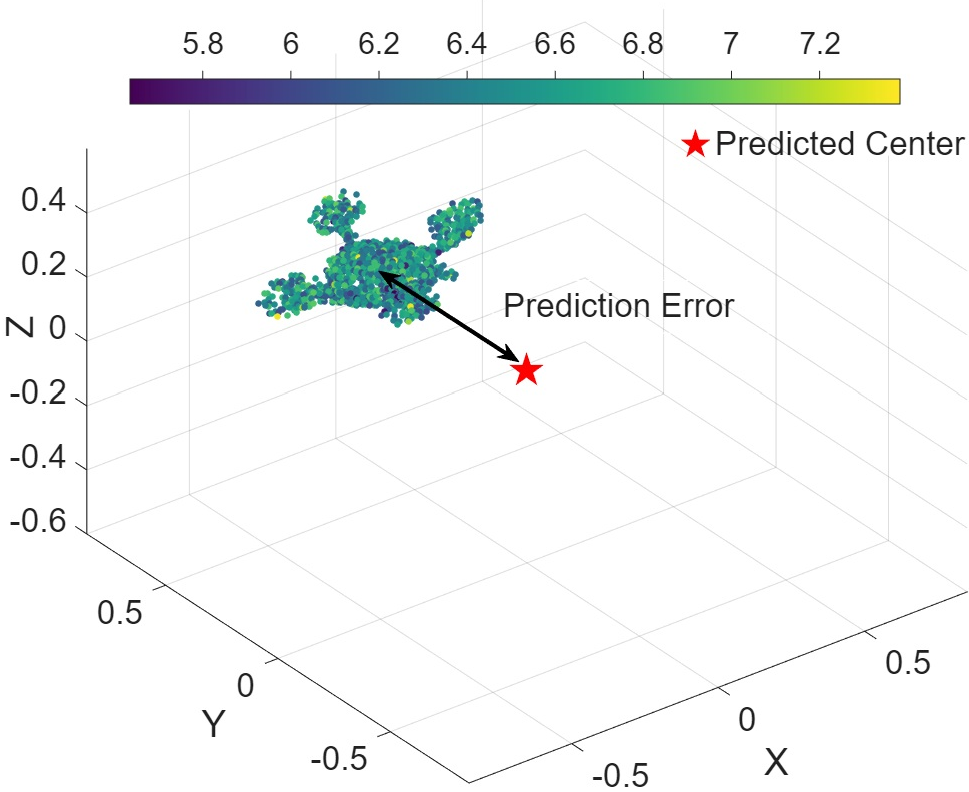}}\\
\subcaptionbox{\footnotesize Ground Truth\label{PositionPre_n3dB_GT}}%
{\includegraphics[width=0.24\textwidth]{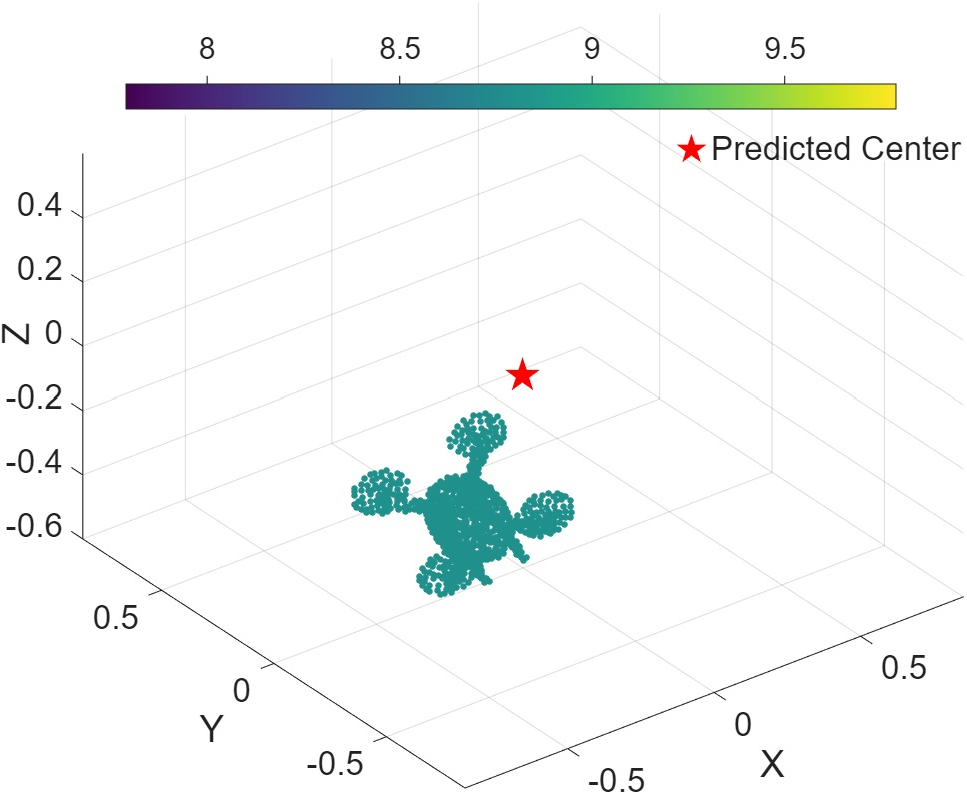}}
\subcaptionbox{\footnotesize Reconstruction ($\text{SNR}\approx -3$ dB)\label{PositionPre_n3dB}}%
{\includegraphics[width=0.24\textwidth]{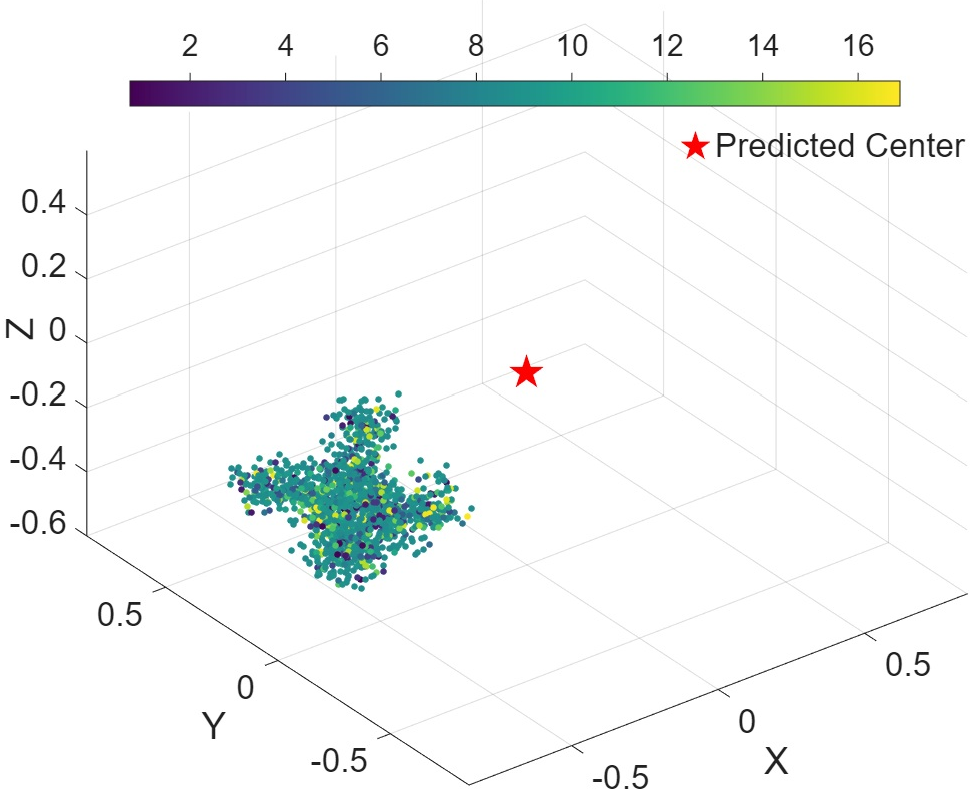}}
\caption{The potential region view of the point cloud reconstruction results based on the AUGUST approach, where the conductivity is presented and the antenna configuration is $16\times4$.}
\label{PositionPre}
\end{figure}

For the diffusion model, 12 ConcatSquash layers of the noise estimation network are structured as 5-16-64-128-256-512-1024-512-256-128-64-16-5. The noise intensity parameter $\beta_s$ linearly increases from 0.0001 to 0.05 over $S=200$ diffusion steps. Regarding the training configurations, the MLP consists of 6 layers, and the other fully-connected networks contain 2 layers. The batch size and training epochs are respectively set to 256 and 200, with a linearly decaying learning rate from $10^{-4}$ to $10^{-5}$. The Adam optimizer is used for optimization. Since we focus more on sensing for the UAV position, attitude, and shape, the weighted coefficients in \eqref{loss concatquash weighted} are set as $\gamma_{\text{pos}}=0.9$ and $\gamma_{\text{EP}}=0.1$. The remaining training parameters are as follows: $d_v = 256$, $d_f = 512$, $d_z = 256$, $d_{\xi} =10$, and $\gamma_{\boldsymbol{z}}=10^{-4}$.

\begin{figure}[t]
    \centering
  \includegraphics[width=.95\linewidth]{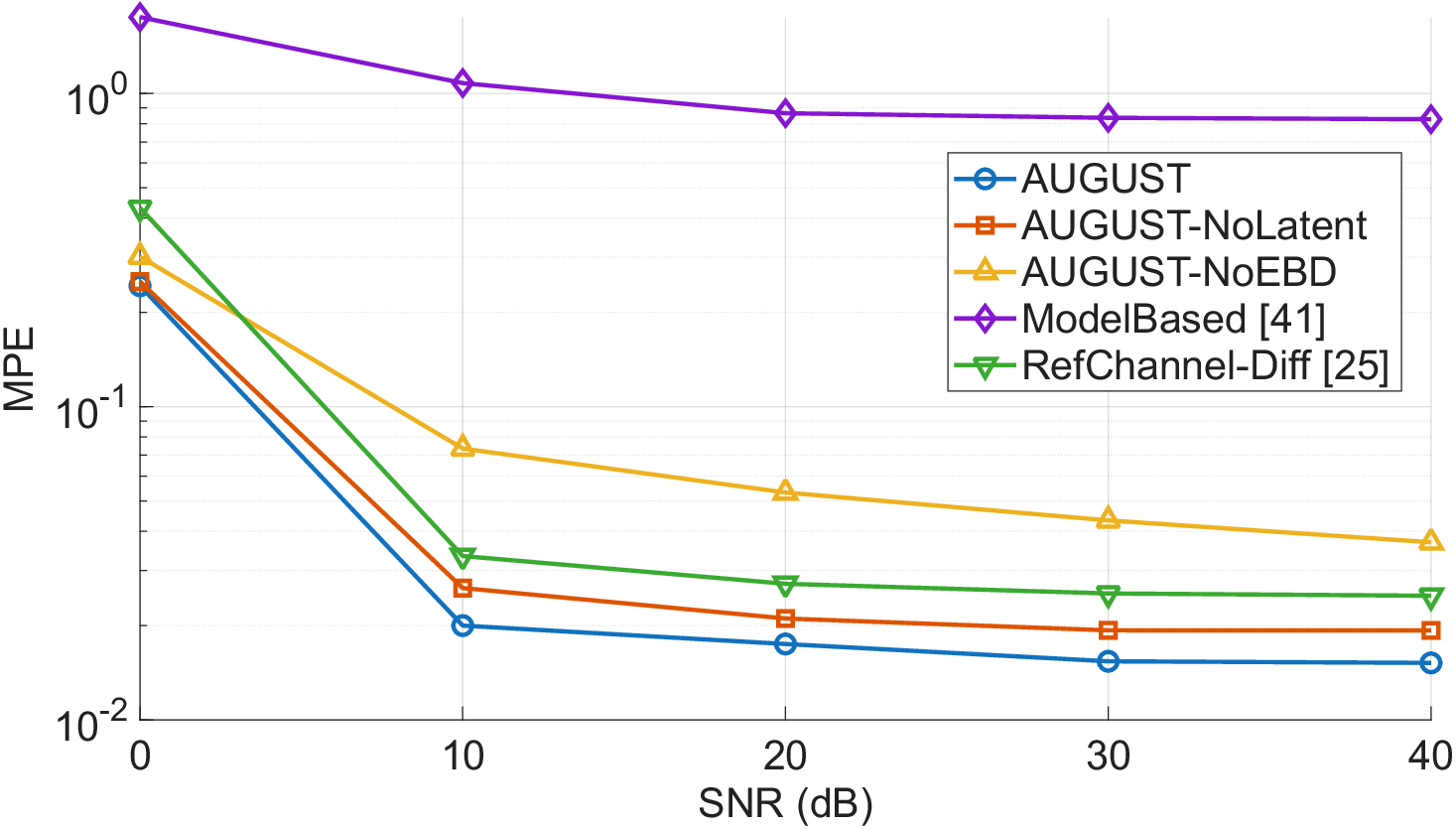}
    \caption{The MPE comparison of the considered approaches under varying SNR conditions, where the antenna configuration is $16\times4$.}
    \label{PositionMSE}
\end{figure}
\begin{figure*}[t]
\centering
\subcaptionbox{\footnotesize Ground Truth\label{AUGUST_GT}}%
{\includegraphics[width=0.195\textwidth]{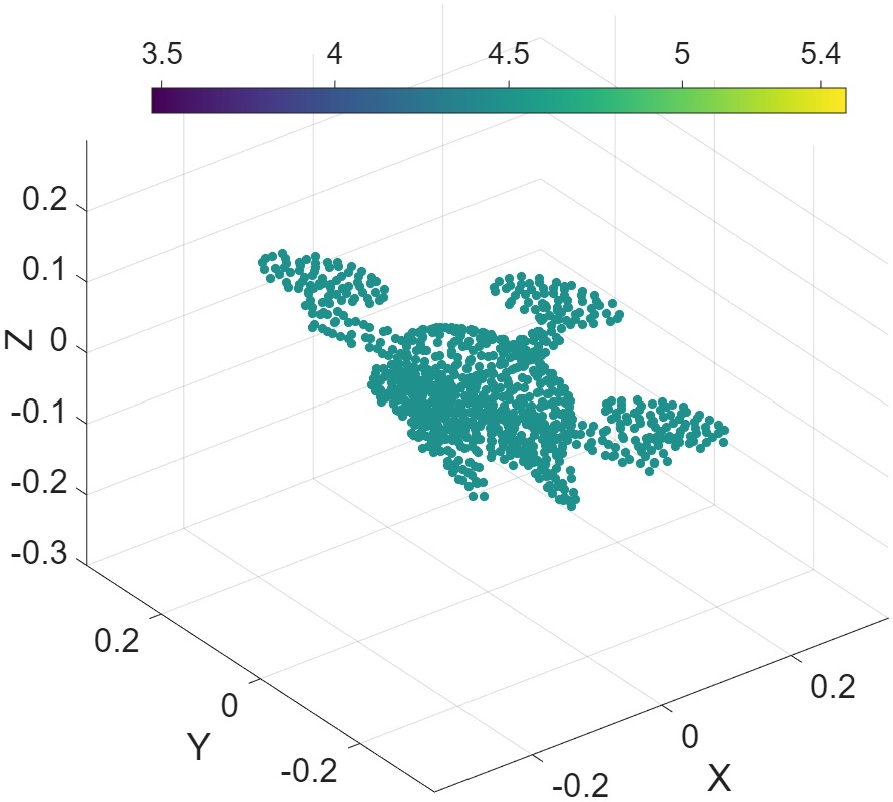}}
\subcaptionbox{\footnotesize AUGUST\label{AUGUST-PC}}%
{\includegraphics[width=0.195\textwidth]{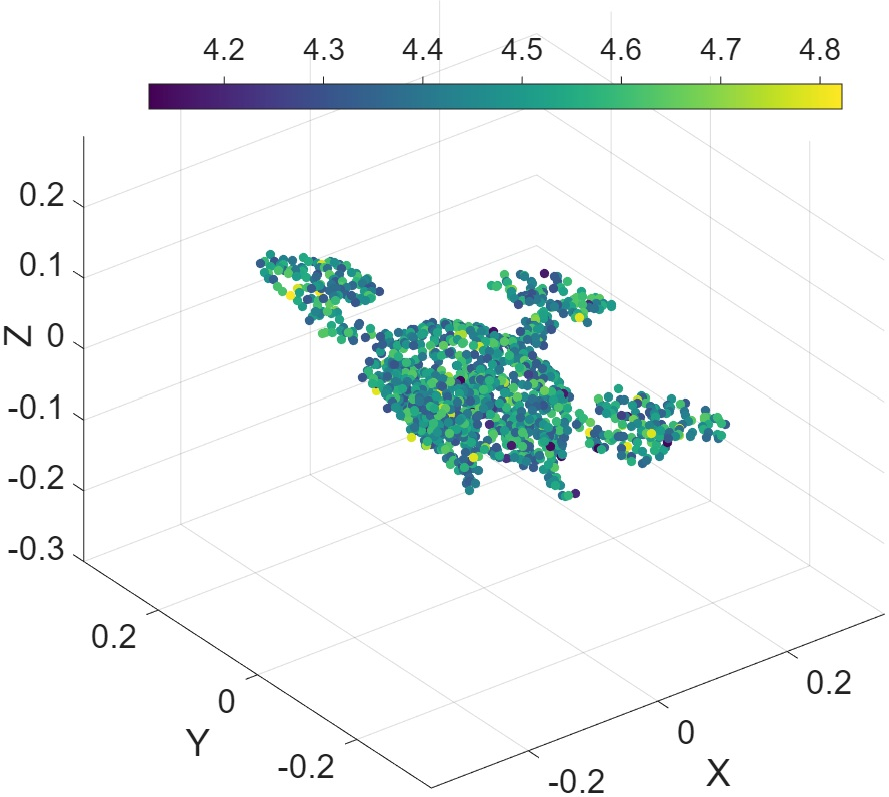}}
\subcaptionbox{\footnotesize AUGUST-NoLatent\label{AUGUST_NoLatent}}%
{\includegraphics[width=0.195\textwidth]{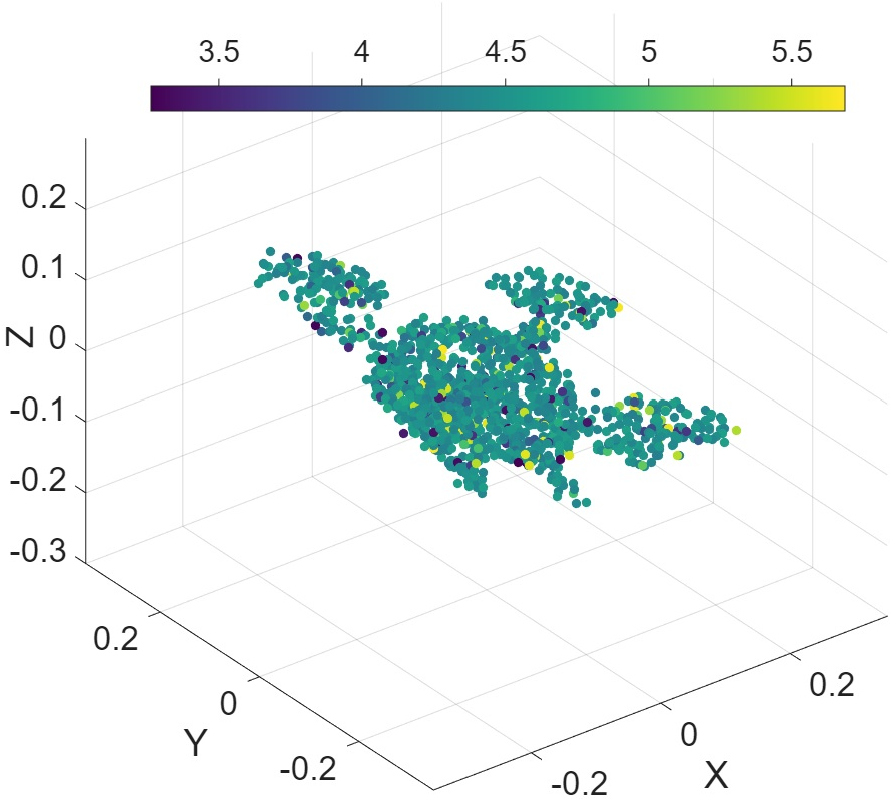}}
\subcaptionbox{\footnotesize AUGUST-NoEBD\label{AUGUST_NoEBD}}%
{\includegraphics[width=0.195\textwidth]{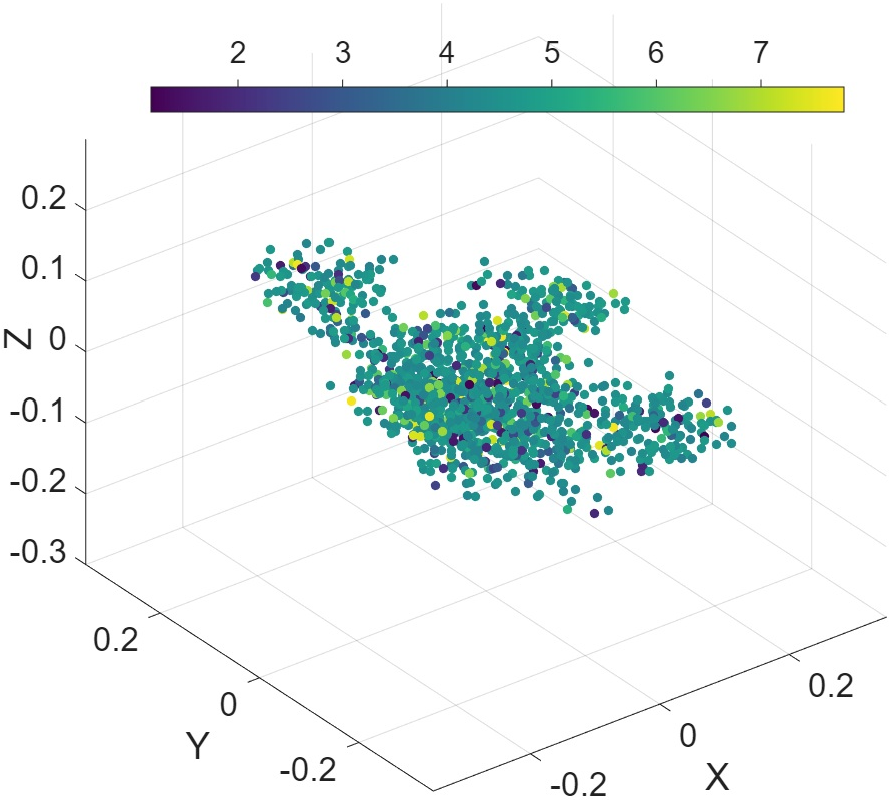}}
\subcaptionbox{\footnotesize RefChannel-Diff\label{RefChannel_Diff}}%
{\includegraphics[width=0.195\textwidth]{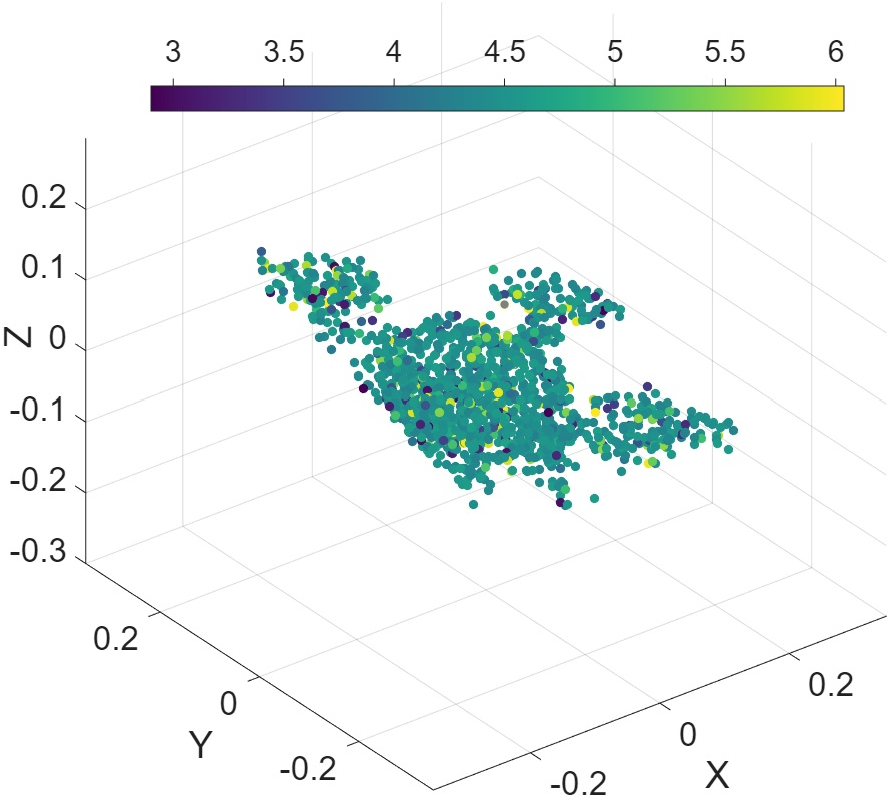}}
\caption{Centralized view of the point cloud reconstruction results based on different approaches, where the relative permittivity is presented, the antenna configuration is $16\times2$, and $\text{SNR}\approx 20$ dB.}
\label{DiffScheme}
\end{figure*}
\begin{figure}[t]
    \centering
  \includegraphics[width=.95\linewidth]{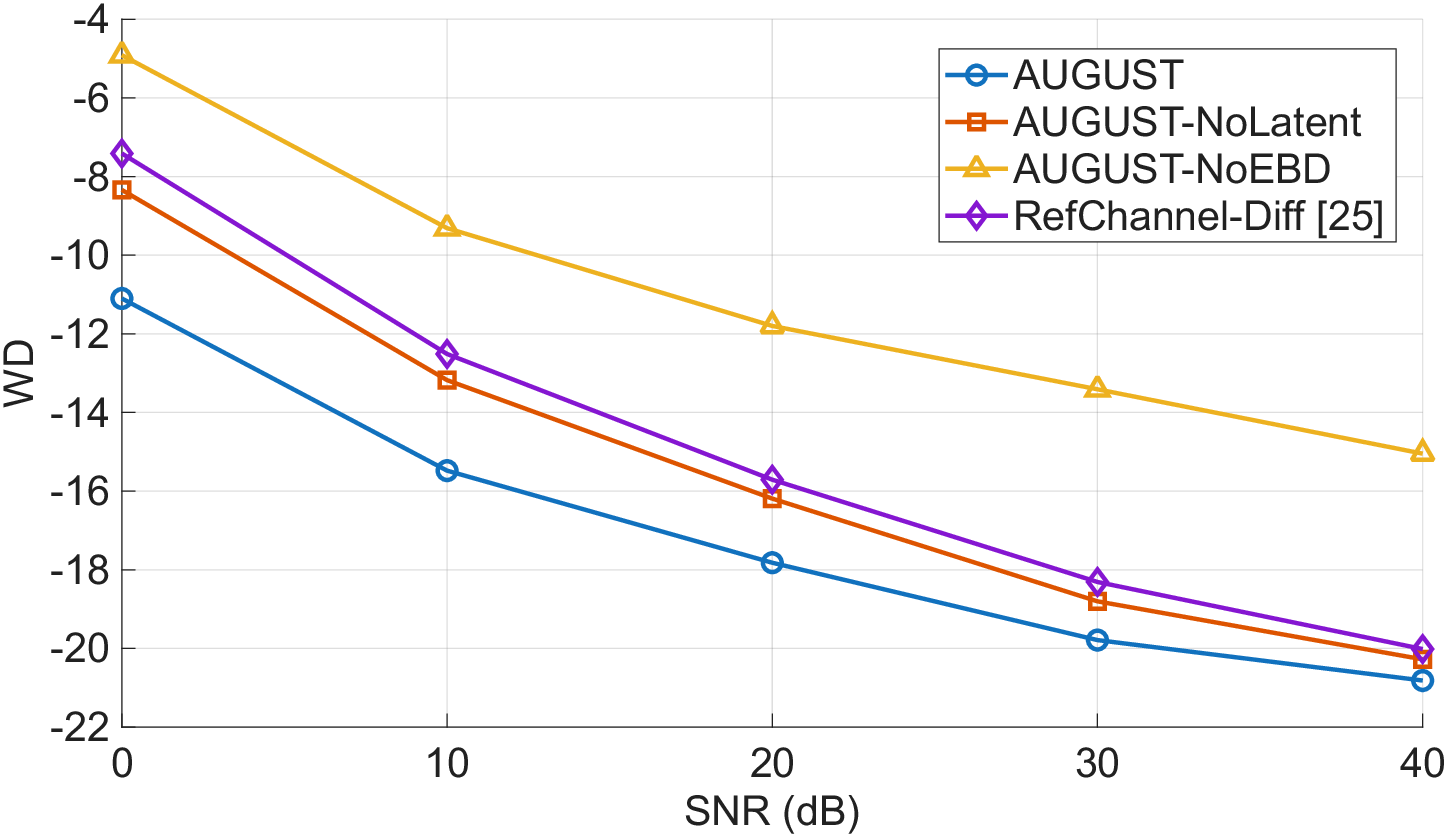}
    \caption{The WD comparison of different approaches under varying SNR conditions, where the antenna configuration is $16\times2$.}
    \label{WD-DiffScheme}
\end{figure}

To evaluate the reconstruction performance of the proposed approach, we adopt an evaluation metric, named as weighted distance (WD), that combines the mean Chamfer distance (MCD) and the MSE of the position estimation:
\begin{align}
    \label{WD}
   & \text{WD}(\text{dB}) = 10\log_{10}\bigg[\frac{1}{T}\sum_{t=1}^{T}\bigg(\frac{1}{M}\sum_{\hat{\boldsymbol{p}}\in\hat{\mathcal{P}}_c^{(t;0)}}\min_{\boldsymbol{p}\in\mathcal{P}_c^{(t;0)}}\|\hat{\boldsymbol{p}}-\boldsymbol{p}\|^2\nonumber\\
    &+ \frac{1}{M}\sum_{\boldsymbol{p}\in\mathcal{P}_c^{(t;0)}}\min_{\hat{\boldsymbol{p}}\in\hat{\mathcal{P}}_c^{(t;0)}}\|\boldsymbol{p}-\hat{\boldsymbol{p}}\|^2+\|\boldsymbol{q}^{(t)}-\boldsymbol{q}_{\mathsf{est}}^{(t)}\|^2\bigg)\bigg],
\end{align}
where $\mathcal{P}_c^{(t;0)}$ and $\hat{\mathcal{P}}_c^{(t;0)}$ denote the centralized UAV point clouds (with the same geometric center) of the ground truth and reconstruction result, respectively; $\boldsymbol{q}^{(t)}$ and $\boldsymbol{q}_{\mathsf{est}}^{(t)}$ correspond to the true and the estimated positions of the UAV based on the geometric centroid of the point cloud 3D coordinates. The first two terms in \eqref{WD} correspond to the MCD that evaluates the reconstruction performance of the point cloud itself, and the remaining term is used to evaluate the accuracy of the UAV position estimation. In addition, to estimate the UAV attitude, we develop a heuristic attitude estimation method motivated by the fact that the UAV rotor plane is parallel to its fuselage plane. Specifically, we denote the partial points on the UAV rotor plane by the point set $\mathcal{S}^{(t)}_{\text{rotor}}$, and define the normalized normal vector of the UAV fuselage plane as $\boldsymbol{\varpi}^{(t)}\triangleq[\varpi_{\rm x},\varpi_{\rm y},\varpi_{\rm z}]^{\mathsf{T}}$. Then, we perform $J$ samplings by randomly selecting three points $\boldsymbol{p}^{(t)}_{a;\jmath}, \boldsymbol{p}^{(t)}_{b;\jmath}, \boldsymbol{p}^{(t)}_{c;\jmath}\in \mathcal{S}_{\text{rotor}}$ and the estimated normal vector $\boldsymbol{\varpi}^{(t)}_{\text{est}}$ can be obtained as:
\begin{align}
    \label{normal vector}
    \boldsymbol{\varpi}^{(t)}_{\text{est}} = \frac{1}{J}\sum^{J}_{\jmath=1}\boldsymbol{\varpi}^{(t)}_{\jmath} = \frac{1}{J}\sum^{J}_{\jmath=1}\frac{(\boldsymbol{p}^{(t)}_{b;\jmath}-\boldsymbol{p}^{(t)}_{a;\jmath})\times(\boldsymbol{p}^{(t)}_{c;\jmath}-\boldsymbol{p}^{(t)}_{a;\jmath})}{\|(\boldsymbol{p}^{(t)}_{b;\jmath}-\boldsymbol{p}^{(t)}_{a;\jmath})\times(\boldsymbol{p}^{(t)}_{c;\jmath}-\boldsymbol{p}^{(t)}_{a;\jmath})\|}.
\end{align}
Based on the cosine similarity, we consider a mean directional error (MDE) to evaluate the performance of UAV attitude (normal vector) estimation, which is defined as $\text{MDE} = \frac{1}{T}\sum_{t=1}^{T}[1-(\boldsymbol{\varpi}^{(t)})^{\mathsf{T}}\boldsymbol{\varpi}^{(t)}_{\text{est}}]$. All inference processes are performed on the NVIDIA RTX 3090 platform, and the average inference time for each point cloud reconstruction is approximately 0.17 seconds when $M=1000$.

For performance comparison, we consider the following benchmark schemes: \textbf{1)} A simplified AUGUST approach that removes the position and SNR embedding (denoted as AUGUST-NoEBD); \textbf{2)}
A simplified AUGUST approach that ignores the mapping of the latent space and the flow prior regularization (denoted as AUGUST-NoLatent), in which the feature $\boldsymbol{f}^{(t)}$ is directly used as the conditional input to the reconstruction module; \textbf{3)} The approach proposed in \cite{jiang2024electromagnetic}, which transforms all estimated sensing channels to a reference channel at a fixed location through positional encoding and then uses the transformed channel as the conditional information for the diffusion model (denoted as RefChannel-Diff). In addition, to compare the performance of UAV position estimation at different slots, we further consider a model-based benchmark for angular and distance estimation as referred to \cite{cui2022channel}, and the mean positioning error (MPE), i.e., $\text{MPE} = \frac{1}{T}\sum_{t=1}^{T}\|\boldsymbol{q}^{(t)}-\boldsymbol{q}_{\mathsf{est}}^{(t)}\|$, is used as an evaluation metric.

\subsection{Performance of Position Estimation}
{\color{blue}To clearly reveal the working mechanism of the proposed AUGUST framework, Fig.~\ref{R3 denoise process} visualizes the reverse diffusion process of the EP point cloud reconstruction conditioned on the encoded sensing-channel features. Starting from the random point cloud in Fig.~\ref{R3 step1}, the reverse process gradually drives the points toward the true UAV location. As shown in Fig.~\ref{R3 step80} and Fig.~\ref{R3 step150}, the point cloud first contracts and aggregates around the target position, suggesting that the low-dimensional position-related information is captured earlier than fine-grained shape and EP details during the reverse process. Then, in the later stages shown in Fig.~\ref{R3 step180} and Fig.~\ref{R3 step200}, the point cloud is progressively denoised and refined to recover the UAV shape, attitude, and more accurate EP information.}

Fig. \ref{PositionPre} illustrates the EP point cloud reconstruction results of the proposed AUGUST approach within the predicted flight region under both high- and low-SNR conditions. From Fig. \ref{PositionPre_20dB_GT} and \ref{PositionPre_n3dB_GT}, it can be seen that, due to the prediction error in UAV position, the UAV may lie anywhere within the region centered on the predicted position. As shown in the upper two subfigures of Fig. \ref{PositionPre}, under high-SNR conditions, the AUGUST approach reconstructs the UAV EP point cloud with high fidelity, capturing its relative 3D position, attitude, and shape information within the reconstruction region. In this manner, for each slot during the tracking, the UAV position can be estimated by adding the reconstructed offset to the predicted center. This validates the effectiveness of the proposed AUGUST approach for UAV sensing and tracking. In contrast, as shown in the lower two subfigures of Fig. \ref{PositionPre}, when the SNR is very low, the reconstruction quality degrades markedly, resulting in pronounced positioning errors and making the extraction of the attitude and shape information considerably more challenging. This indicates that low-SNR conditions significantly hinder the network’s ability to extract UAV features from the estimated sensing channel.

Furthermore, Fig. \ref{PositionMSE} presents a comparison of the UAV position estimation performance across different schemes. Clearly, the proposed AUGUST approach achieves significantly lower estimation errors than the other alternatives, reaching centimeter-level accuracy. Moreover, all diffusion-model-based approaches consistently outperform the conventional model-based method in terms of positioning accuracy. In particular, when the channel conditions are moderately improved, i.e., $\text{SNR}\ge 10$ dB, these approaches exhibit notable performance gains. This observation indicates the strong capability of the diffusion model for extracting UAV position-related information from the sensing channel.
\begin{figure*}[htb]
\centering
\subcaptionbox{\footnotesize Ground Truth\label{Diff_Ant_GT}}%
{\includegraphics[width=0.245\textwidth]{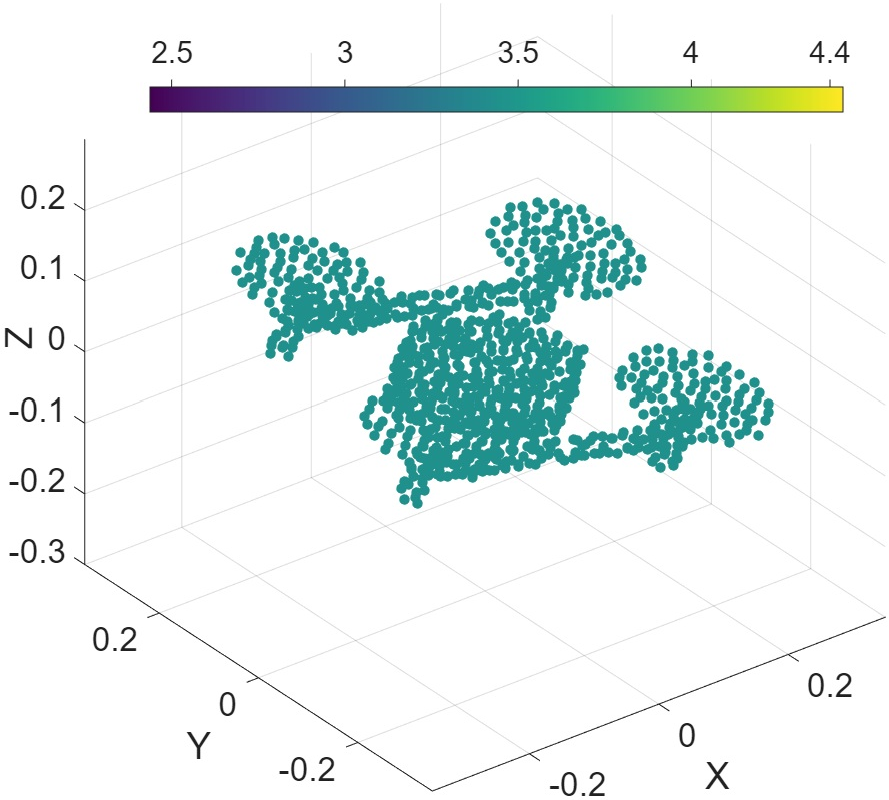}}
\subcaptionbox{\footnotesize $16\times 4$ antennas\label{Diff_Ant_16x4}}%
{\includegraphics[width=0.245\textwidth]{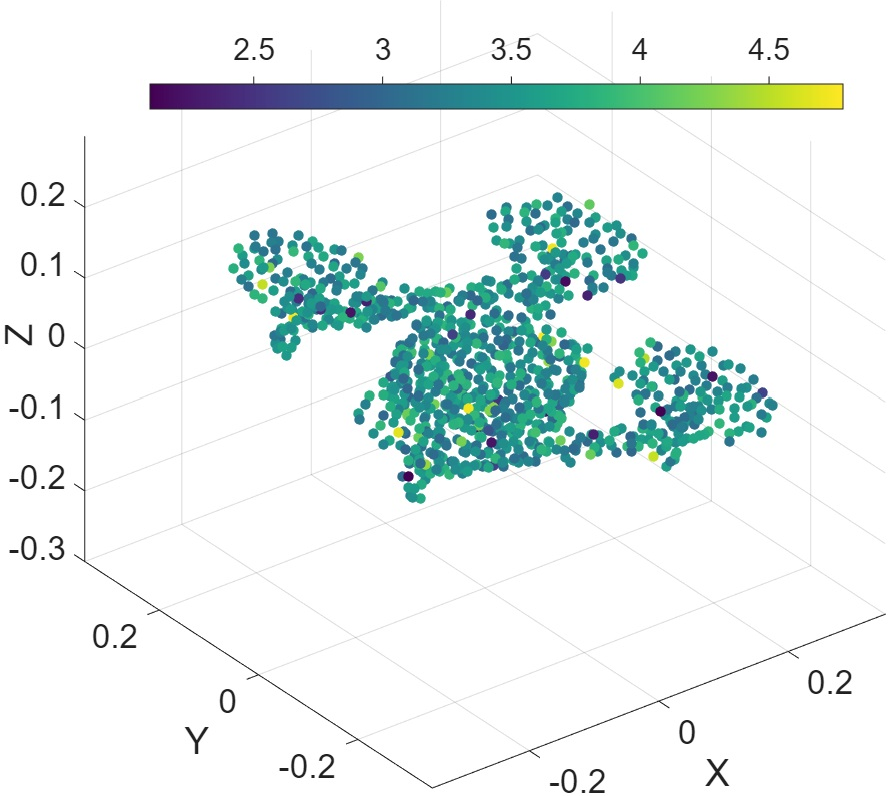}}
\subcaptionbox{\footnotesize $16\times 2$ antennas\label{Diff_Ant_16x2}}%
{\includegraphics[width=0.245\textwidth]{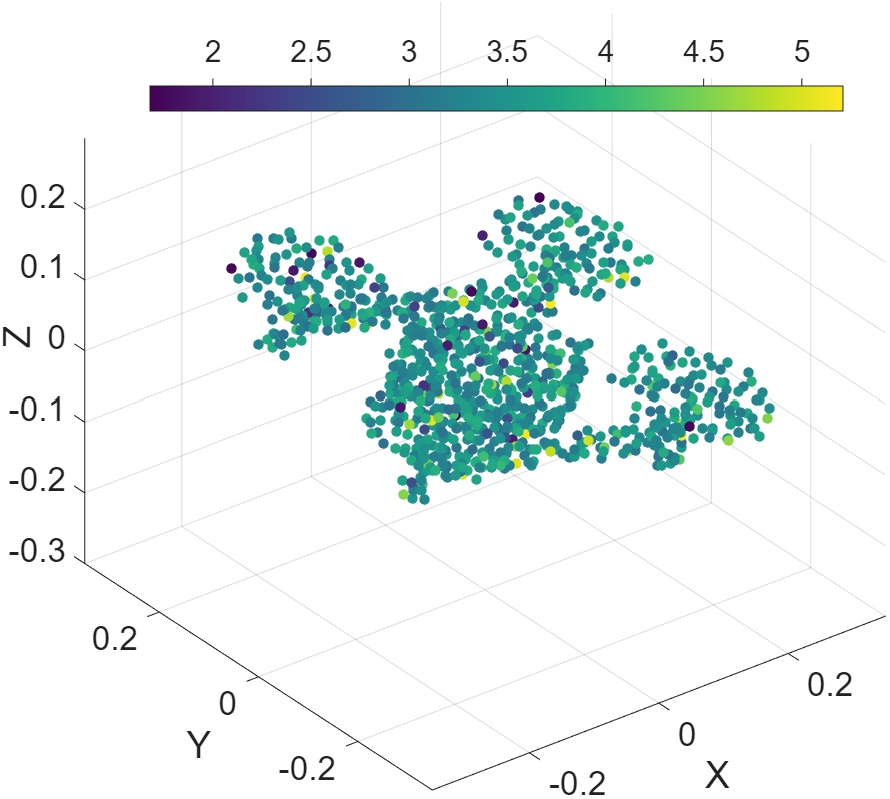}}
\subcaptionbox{\footnotesize $16\times 1$ antennas\label{Diff_Ant_16x1}}%
{\includegraphics[width=0.245\textwidth]{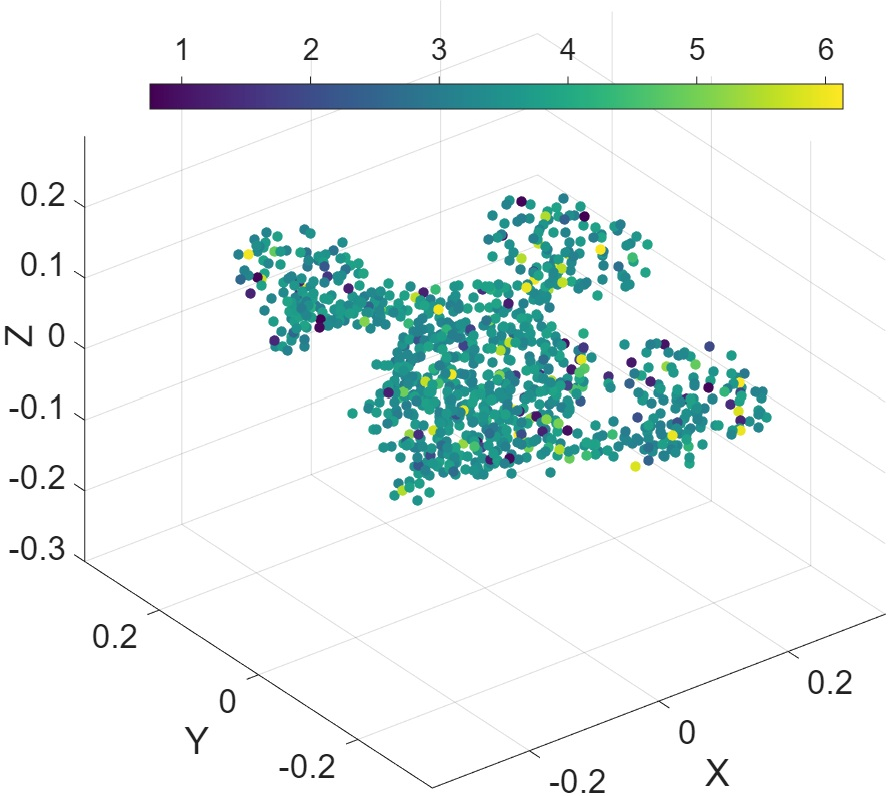}}
\caption{Centralized view of the point cloud reconstruction results based on the AUGUST approach under different antenna configurations, where the conductivity is presented, and $\text{SNR}\approx 20$ dB.}
\label{DiffAnt}
\end{figure*}
\begin{figure}[t]
    \centering
  \includegraphics[width=.95\linewidth]{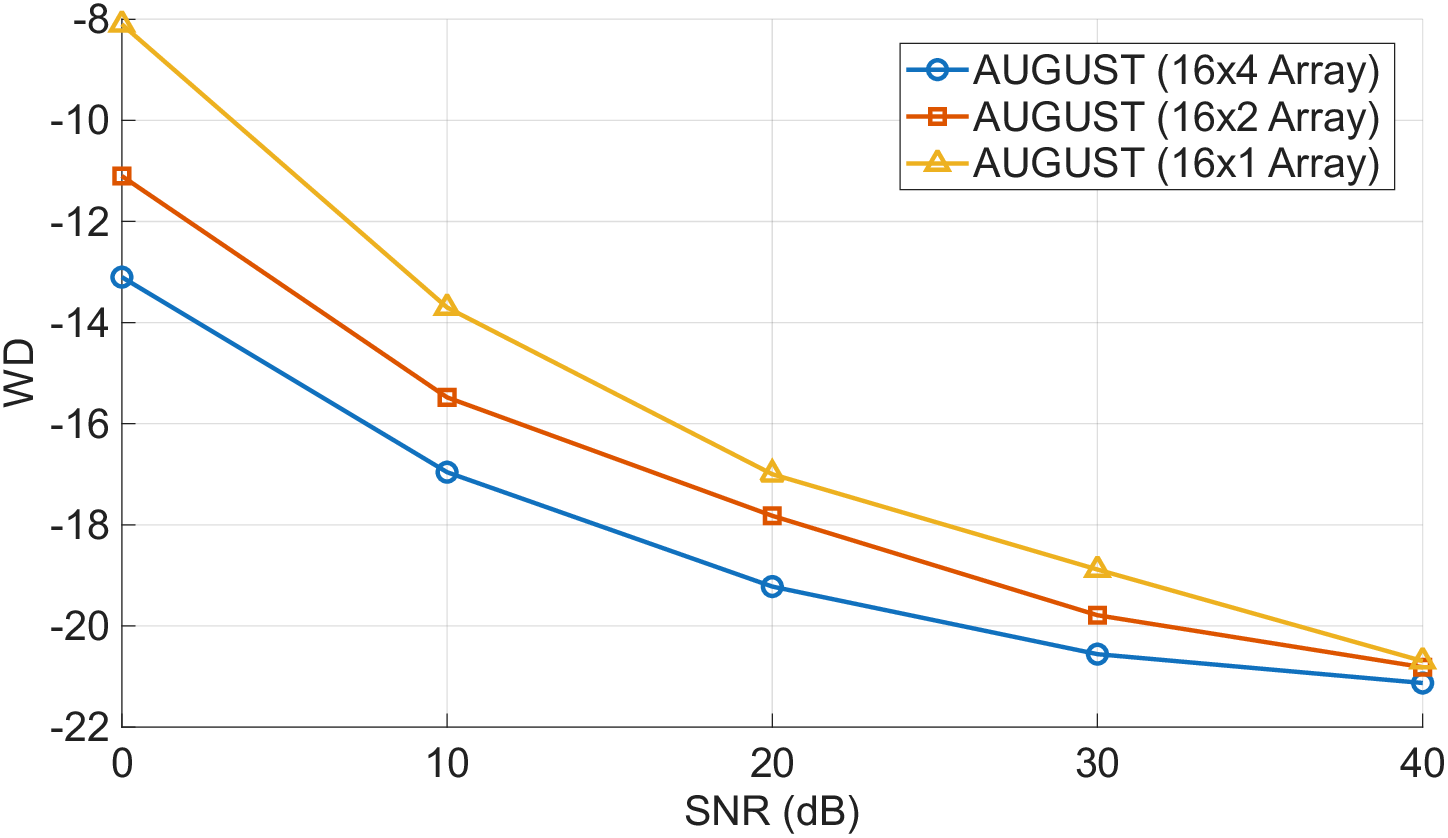}
    \caption{The WD comparison of the AUGUST approach under different antenna configurations and SNR conditions.}
    \label{WD-DiffAnt}
\end{figure}
\subsection{Performance of Point Cloud Reconstruction}
To enable a more intuitive comparison of reconstruction quality across the considered schemes, a centralized-view visualization of the reconstructed point clouds is presented in Fig. \ref{DiffScheme}. It is observed from Fig. \ref{DiffScheme} that the AUGUST approach yields point clouds substantially closer to the ground truth than the benchmark schemes, whose reconstructions exhibit varying degrees of contamination by noisy points. In particular, the reconstruction quality of the AUGUST-NoLatent and RefChannel-Diff approaches is comparable, both exhibiting a small number of noisy points. This can be attributed to the fact that both schemes rely on positional encoding to obtain deterministic conditioning information from the sensing channel, either via feature extraction or channel mapping. Specifically, the former extracts the channel features into a deterministic feature vector, whereas the latter maps the estimated channel to a reference channel, which then serves as the conditional information for diffusion denoising. In contrast, AUGUST-NoEBD yields the poorest reconstruction, with the resulting point cloud being severely corrupted. This is because the AUGUST-NoEBD neglects both positional and SNR embeddings, making it difficult for the network to accurately extract intrinsic features of the UAV from the sensing channel in the absence of these auxiliary cues. These results indicate that mapping the extracted channel features into a latent space and sampling from it yield cleaner point cloud reconstructions than directly using deterministic features as inputs. Meanwhile, the necessity of incorporating the position and SNR information is also clearly validated. Correspondingly, Fig. \ref{WD-DiffScheme} quantifies the differences in Fig. \ref{DiffScheme} in terms of WD and confirms that the proposed AUGUST approach achieves the lowest WD, i.e., the best reconstruction quality.

\begin{figure}[t]
    \centering
  \includegraphics[width=.95\linewidth]{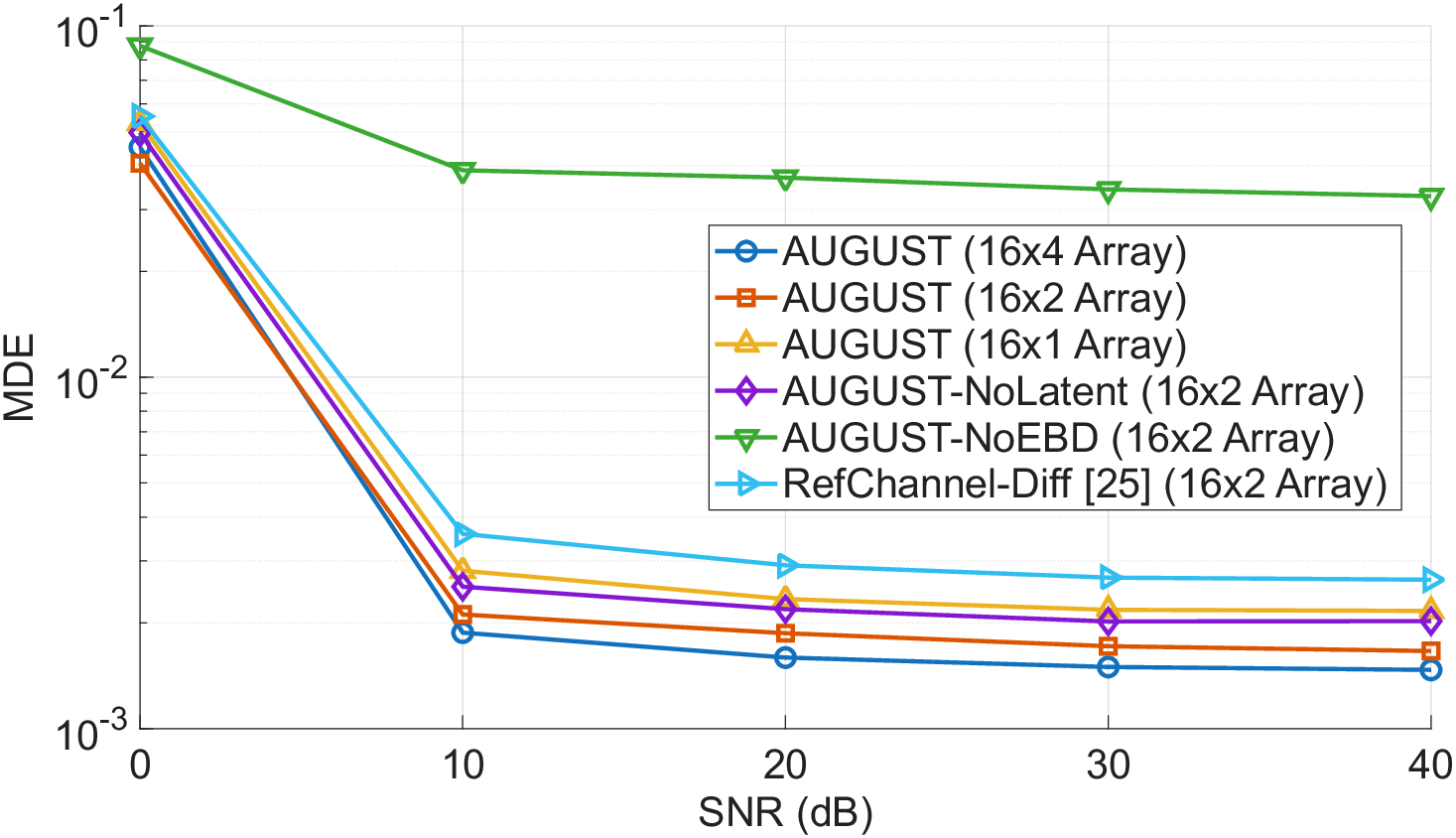}
    \caption{The MDE comparison of the considered approaches under varying SNR conditions and antenna configurations.}
    \label{DE}
\end{figure}

As one of the key factors influencing sensing performance, the number of antennas directly determines the dimension of the sensing channel, i.e., the dimension of the observed data. To investigate this effect, Figs. \ref{DiffAnt} and \ref{WD-DiffAnt} compare the reconstruction performance of the AUGUST approach under different antenna configurations via imaging results and the quantitative comparison in terms of the WD metric, respectively. Specifically, Fig. \ref{DiffAnt} illustrates the centralized-view point cloud reconstruction results for three antenna setups. It can be clearly observed that the point cloud reconstructed with the $16\times1$ antenna configuration contains more noisy points. As the number of antennas increases, the reconstructed UAV point cloud under the same channel conditions becomes progressively closer to the ground truth. This trend suggests that a higher observation dimensionality facilitates more accurate reconstruction of the EP distribution of the UAV, thereby enabling more precise feature characterization. Accordingly, this conclusion is further supported by the quantitative results in Fig. \ref{WD-DiffAnt}, evaluated in terms of WD. The AUGUST approach with the $16\times 4$ antenna configuration consistently achieves lower WD values across the considered SNR regimes. Moreover, in the low-SNR regime, it exhibits more pronounced performance gains compared to cases with fewer antennas, highlighting the benefits brought by increased observation dimensionality.

\begin{figure}[t]
\centering
\subcaptionbox{\footnotesize Ground Truth\label{RegionSize_GT}}%
{\includegraphics[width=0.24\textwidth]{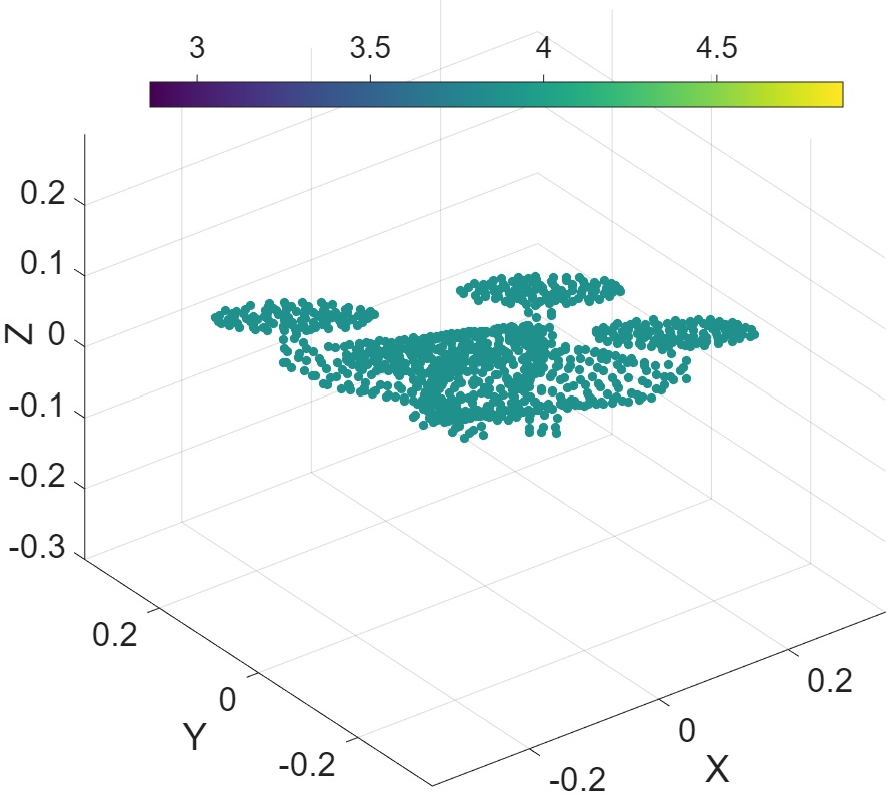}}
\subcaptionbox{\footnotesize $s_{
x/y/z}=0.85$ m\label{RegionSize_85}}%
{\includegraphics[width=0.24\textwidth]{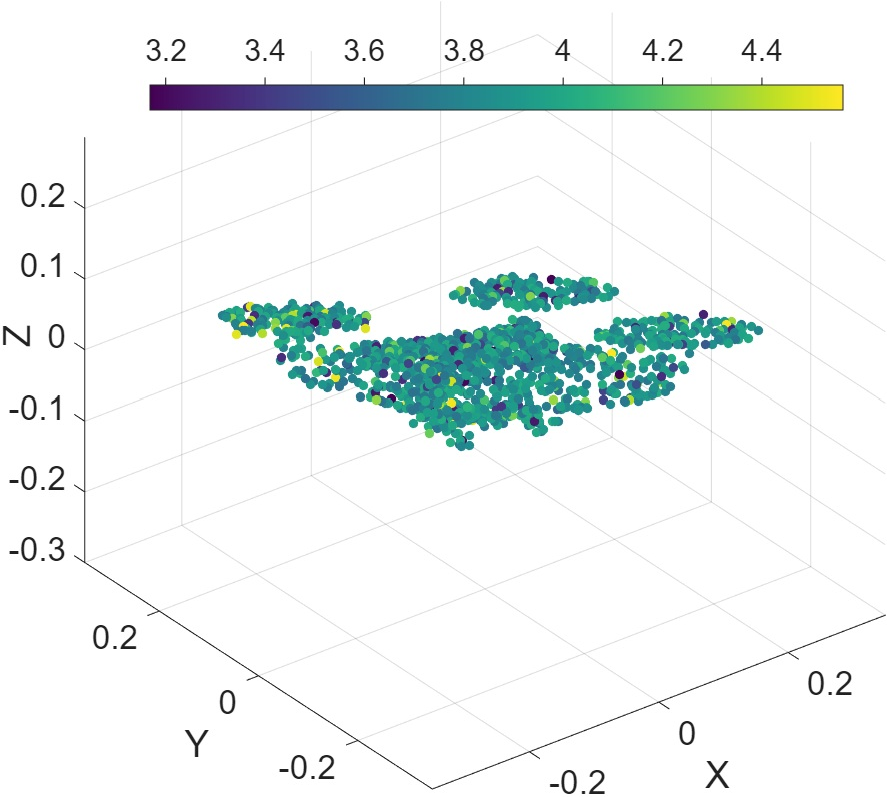}}\\
\subcaptionbox{\footnotesize $s_{
x/y/z}=1.5$ m\label{RegionSize_150}}%
{\includegraphics[width=0.24\textwidth]{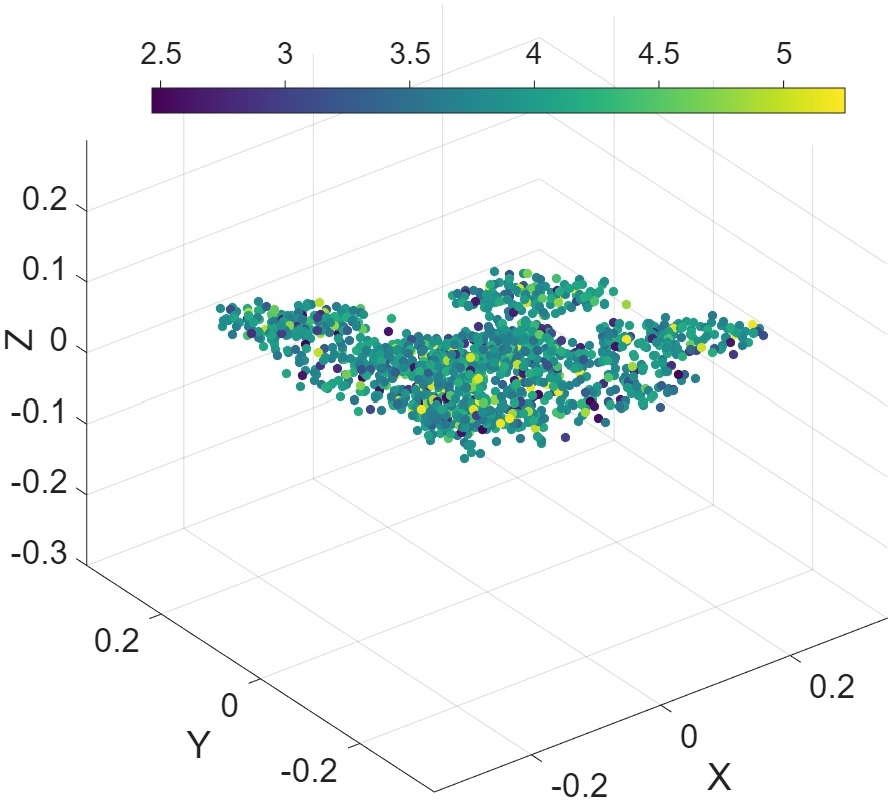}}
\subcaptionbox{\footnotesize $s_{
x/y/z}=2.5$ m\label{RegionSize_250}}%
{\includegraphics[width=0.24\textwidth]{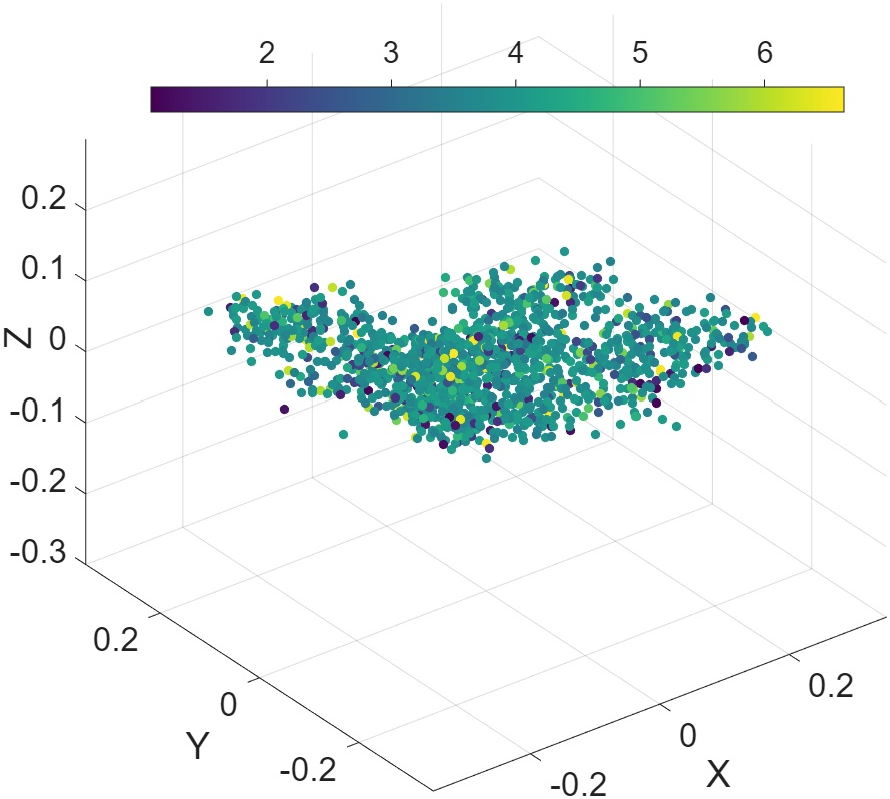}}\\
\subcaptionbox{\footnotesize \color{blue} WD comparison\label{WD-DiffRegionSize}}%
{\includegraphics[width=0.45\textwidth]{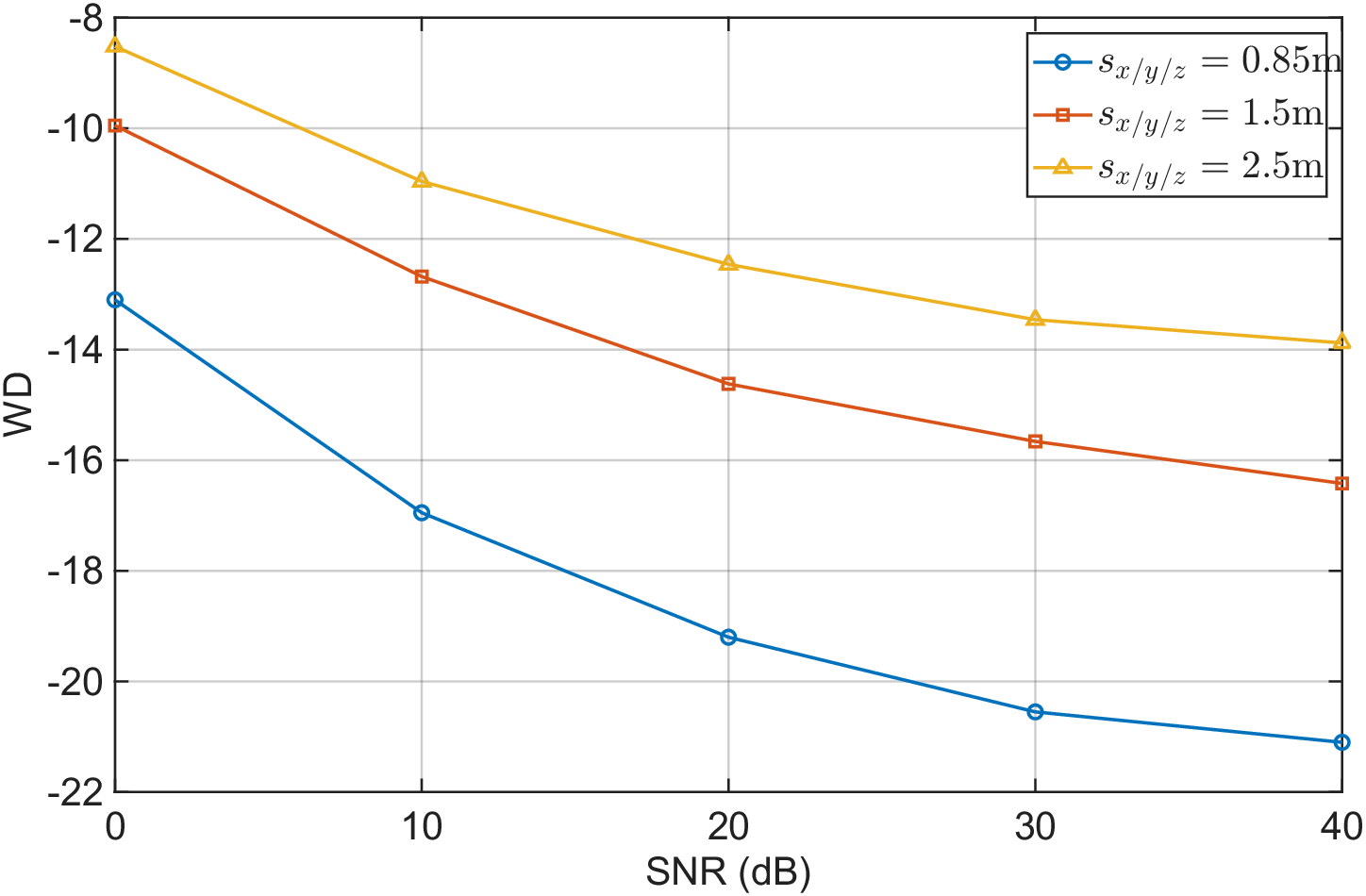}}
\caption{\color{blue}Centralized view of the point cloud reconstruction results and the WD comparison based on the AUGUST approach under different region sizes $s_{
x/y/z}$. For (a)-(d), the relative permittivity is presented, the antenna configuration is $16\times4$, and $\text{SNR}\approx 30$ dB.}
\label{RegionSize}
\end{figure}

From the reconstructed UAV point cloud, the UAV shape is readily discernible. Its attitude, represented by the normal vector of the fuselage plane, can be estimated using the heuristic method in \eqref{normal vector}. Fig. \ref{DE} provides a comprehensive comparison of the MDE achieved by the AUGUST approach and competing schemes under different settings. From Fig. \ref{DE}, it is straightforward to observe that under poor channel conditions (i.e., $\text{SNR}\approx 0$ dB), none of the schemes can accurately estimate the fuselage-plane normal vector, resulting in an MDE close to $10\%$. This is because the previous results have shown that unfavorable channel conditions introduce substantially more noisy points in the reconstructed point cloud, which hinders the reliable capture of attitude information, as illustrated in Fig. \ref{AUGUST_NoEBD}. In addition, Fig. \ref{AUGUST_NoEBD} also offers intuitive evidence for why the AUGUST-NoEBD approach consistently yields higher MDE than the other schemes. By contrast, once the channel condition improves, the considered schemes except AUGUST-NoEBD can achieve a lower MDE, and the performance differences among the different solutions are not significant. Together with the imaging results in Fig. \ref{DiffScheme}, these results suggest that the heuristic attitude estimator in \eqref{normal vector} remains effective for point clouds with a moderate level of contamination, yielding stable normal-vector estimation. This aligns with the visual intuition that the UAV attitude and shape remain discernible from mildly corrupted images. Moreover, the results in Fig. \ref{DE} further confirm that point cloud imaging under the GAI framework can effectively capture the UAV attitude information.

    \begin{figure}[t]
        \centering
        \subcaptionbox{\footnotesize \color{blue} Smooth case with maximum prediction error 0.41 m and $s_{x/y/z}=0.85$m \label{R2 Traj 50cm C4}}%
        {\includegraphics[width=0.49\textwidth]{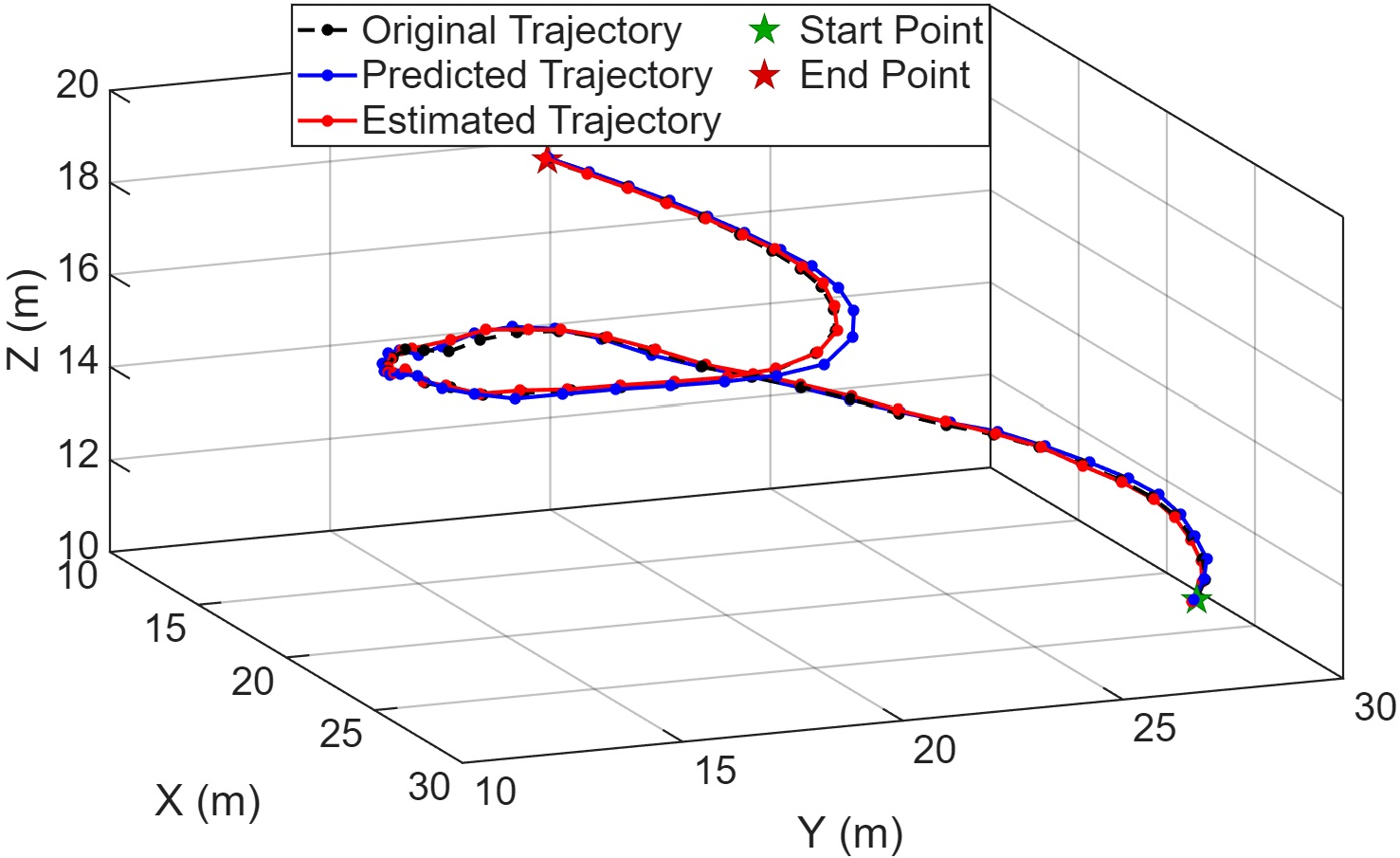}}
        \subcaptionbox{\footnotesize \color{blue} Dynamic case with maximum prediction error 1.93 m and $s_{x/y/z}=2.5$m \label{R2 Traj 250cm C4}}%
        {\includegraphics[width=0.49\textwidth]{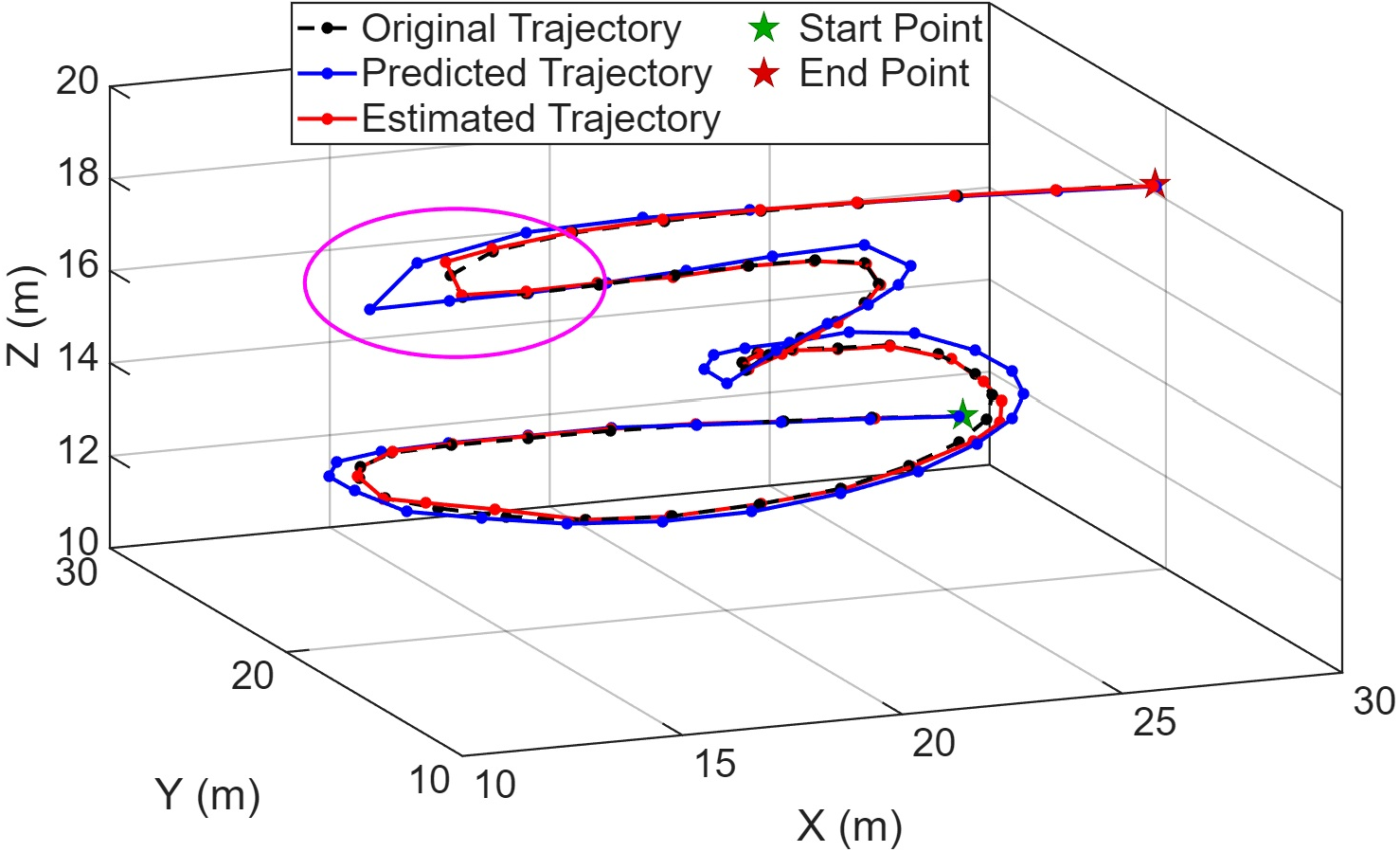}}
        \caption{\color{blue}Prediction and tracking results for two samples of UAV trajectories generated with varying randomness.}
        \label{R2 Traj C4}
    \end{figure}

\subsection{Additional Discussions on Key Parameters}
As defined in \eqref{5D pc}, the size of the reconstruction region, i.e., $s_{x/y/z}$, hinges on the UAV size and the position prediction error threshold. This means that a larger prediction error threshold requires point cloud reconstruction over a larger region, corresponding to a weaker spatial prior for UAV tracking in each slot. Fig. \ref{RegionSize} briefly examines the impact of region size by showing the point cloud reconstruction results of the AUGUST approach for a fixed-size UAV under different $s_{x/y/z}$. It is observed from Fig. \ref{RegionSize} that, as $s_{x/y/z}$ increases, the reconstructed point cloud becomes increasingly corrupted. In particular, when $s_{x/y/z}=2.5$ m, the UAV features become difficult to discern. {\color{blue}To further analyze the impact of the selection of $s_{x/y/z}$ quantitatively, we additionally compare the WD performance of AUGUST under different $s_{x/y/z}$, as shown in Fig.~\ref{WD-DiffRegionSize}. It can be observed that, under the same SNR, a smaller $s_{x/y/z}$ consistently achieves a lower WD. This is consistent with the conclusions drawn from the point cloud imaging results above.} {\color{blue}The above results also reflect the influence of prediction uncertainty on current-slot sensing performance. A larger $s_{x/y/z}$ can tolerate larger errors in $\boldsymbol{q}^{(t)}_{\mathsf{pre}}$, including those caused by historical position estimation errors and trajectory randomness. However, for a fixed-size UAV, enlarging the reconstruction region makes the UAV occupy a smaller relative portion of the region and dilutes the UAV EP distribution over a larger space. Therefore, there exists a trade-off between robustness to prediction errors and detailed point cloud reconstruction quality. In practice, $s_{x/y/z}$ should be selected according to the expected prediction error level and the desired reconstruction accuracy.}

{\color{blue}To intuitively evaluate the performance of UAV position prediction and tracking, we compare two UAV trajectories with different motion patterns. As shown in Fig.~\ref{R2 Traj 50cm C4}, for the smooth trajectory, the predicted trajectory follows the original trajectory well, with a maximum prediction error of $0.41$ m. Accordingly, given a region size $s_{x/y/z}=0.85$ m that can cover the UAV during the tracking, the estimated trajectory obtained by AUGUST closely matches the original trajectory. For the dynamic trajectory in Fig.~\ref{R2 Traj 250cm C4}, the UAV changes its moving direction more frequently, especially around the sharp maneuvering region, which increases the maximum prediction error to $1.93$ m. In this case, we need to enlarge the reconstruction region to $s_{x/y/z}=2.5$ m to ensure that the UAV remains covered by the potential region. Although the predicted trajectory shows noticeable deviations in the high-maneuvering segment, the estimated trajectory extracted from the reconstructed point clouds still follows the original trajectory well for most slots. These results indicate that trajectory randomness mainly affects the prediction accuracy of $\boldsymbol{q}^{(t)}_{\mathsf{pre}}$, while the final tracking performance of AUGUST can remain reliable when the potential flight region is properly selected. This is because $\boldsymbol{q}^{(t)}_{\mathsf{pre}}$ only determines the region center, whereas the current-slot position $\boldsymbol{q}^{(t)}_{\mathsf{est}}$ is estimated from $\hat{\mathcal{P}}^{(t)}$. Therefore, historical trajectory estimation errors and position prediction errors do not simply accumulate in an open-loop manner, and can be mitigated or corrected by current-slot sensing and imaging.}

\begin{figure}[t]
                \centering
                \subcaptionbox{\footnotesize \color{blue} WD comparison for varying UAV size settings.
                \label{R3 diff-uavsize}}[0.95\linewidth]{\includegraphics[width=\linewidth]{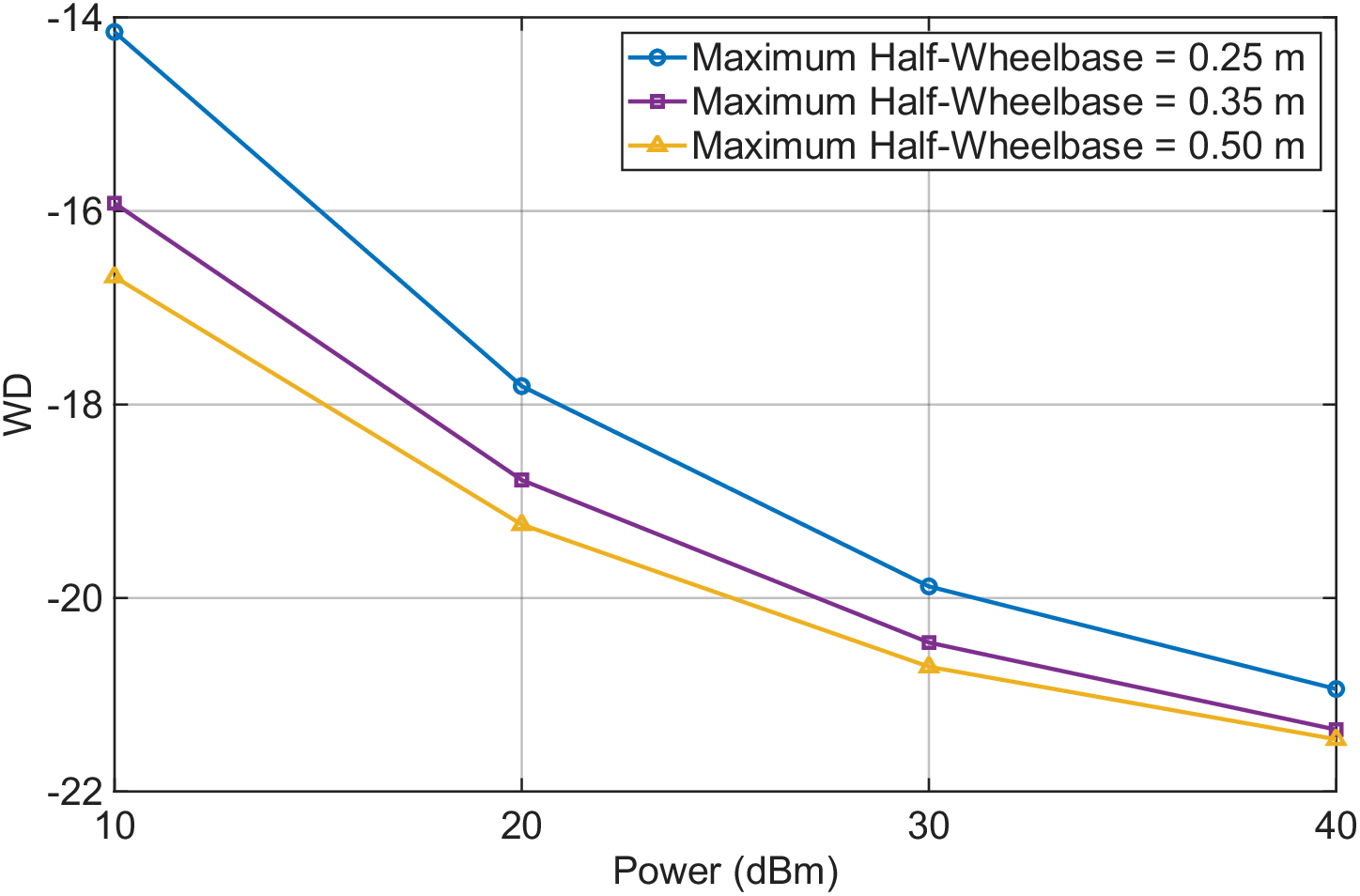}}\\
                
                \subcaptionbox{\footnotesize \color{blue} WD comparison for varying material settings.
                \label{R3 diff-material}}[0.95\linewidth]{\includegraphics[width=\linewidth]{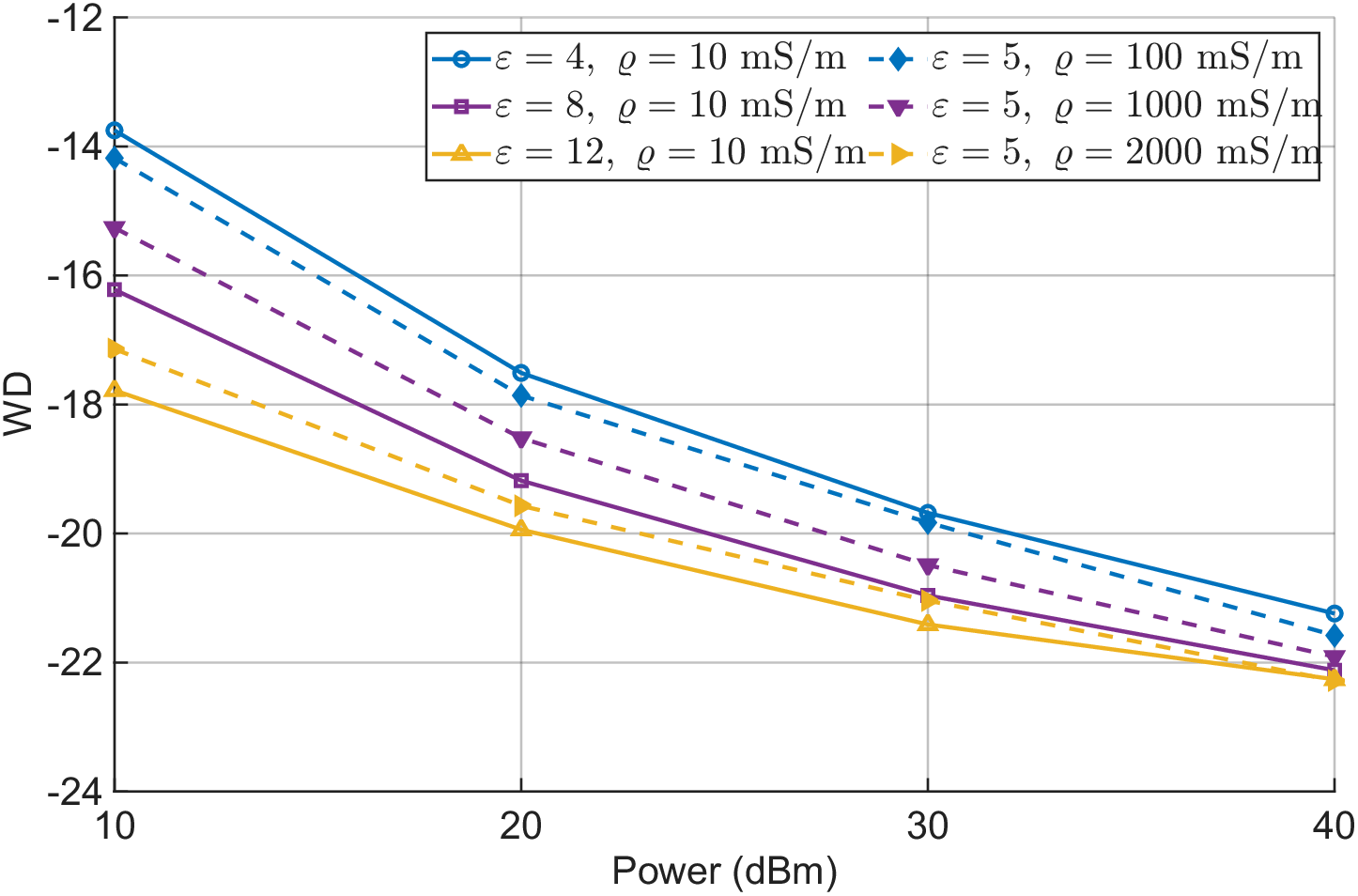}}
                \caption{\color{blue}The WD comparisons based on AUGUST approach under different UAV size and material settings, where the antenna configuration is $16\times4$.}
                \label{R3 diff-uavsize-material}
            \end{figure}

{\color{blue}To further evaluate the scalability of the proposed AUGUST approach, we consider more diverse UAV size and material settings. The corresponding datasets are regenerated and the} {\color{blue}proposed model is retrained accordingly. For the UAV size setting, the maximum half-wheelbase (i.e., half of the maximum diagonal distance between two opposite rotors) is varied from $0.25$ m to $0.35$ m and $0.50$ m, while the other UAV components are proportionally scaled. As shown in Fig.~\ref{R3 diff-uavsize}, a larger UAV size generally leads to a lower WD under the same transmit power. This is because a larger UAV usually has a larger effective scattering aperture and angular extent observed by the antenna array, making the UAV-induced spatial features more distinguishable in the sensing channel and thereby improving point cloud reconstruction. For the material setting, the relative permittivity and conductivity are extended to $\varepsilon\in[1.5,15]$ and $\varrho\in[10,2000]$ mS/m, respectively, and the corresponding dataset is regenerated for training and testing. As shown in Fig.~\ref{R3 diff-material}, the WD decreases as the transmit power increases, indicating that a higher received SNR leads to more accurate point cloud reconstruction. Moreover, under the same transmit power, larger $\varepsilon$ or $\varrho$ generally yields a lower WD. This is because stronger permittivity contrast and conductivity response enhance the target-related scattered echoes under the considered settings, thereby improving the observability of the UAV EP distribution from the sensing channel.}

\section{Conclusion} 
Based on electromagnetic scattering modeling, we leveraged the estimated sensing channel and developed an AUGUST approach to achieve UAV sensing and tracking via EP point cloud imaging within a potential flight region, thereby capturing its position, attitude, and shape information. Our AUGUST approach comprises a channel encoding module and a generative decoding module. The multiplicative position embedding and the SNR embedding are utilized in the encoding module to assist the MLP-based encoder in extracting the stable UAV features under varying UAV locations and channel conditions. The encoded features are further mapped into a latent space that is regularized by a learnable flow-prior. Conditioned on the extracted features, the decoding module adopts a diffusion model to reconstruct the EP point cloud of the UAV based on the weighted training design. Numerical results demonstrate that the proposed AUGUST approach presents higher-fidelity point cloud imaging than the competing schemes, thereby enabling more accurate characterization of the UAV attitude and shape. Moreover, the AUGUST approach achieves markedly better position estimation performance than the conventional positioning method.

{\color{blue}Looking forward, extending the proposed AUGUST approach to more complex propagation environments, such as multipath-rich or blockage-induced NLoS scenarios, constitutes an important direction for future research. In practical urban low-altitude scenarios, static environmental objects, such as buildings, walls, and dominant reflectors, may introduce additional multipath components into the received echoes and may even cause temporary blockage of the UAV-induced LoS echo, resulting in degraded imaging quality or potential track loss. Meanwhile, the EM properties and geometric information of these static environmental objects are often known, calibrated, or slowly varying. Inspired by the diffusion Schrödinger bridge (DSB)-based EM property sensing and channel reconstruction framework proposed in \cite{jiang2025electromagnetic}, such static environmental information could be mapped into environment-aware latent features or reconstructed channels through bidirectional generative modeling. These features could then be incorporated into the conditional channel encoding module of AUGUST as additional environmental priors, helping to distinguish UAV-induced echoes from environment-induced multipath components and to exploit indirect propagation paths in multipath or blockage scenarios. This extension may further improve UAV point cloud imaging, positioning, and tracking robustness in complex low-altitude propagation environments.

Another important research direction is to characterize the effective sensing range of the proposed AUGUST framework in large-scale environments. As indicated by the electromagnetic propagation model in \eqref{Ei} to \eqref{signal yl}, the spatial and electromagnetic-property information of the UAV surface is embedded into the received signal through the propagation of the incident field, the EP-dependent scattering response over the UAV surface, and the projection of the scattered electric field onto the polarization direction of receiving antennas. As the sensing distance increases, the echo SNR decreases and the angular extent of the UAV observed by the finite-aperture antenna array becomes smaller, reducing the distinguishable spatial features carried by the sensing channel. Therefore, the maximum effective sensing range is not a fixed hard boundary determined only by distance, but rather a \textit{``task-dependent soft boundary''} jointly affected by the link budget, array aperture, carrier frequency, UAV size, material properties, attitude, and the required reconstruction accuracy. Such a boundary is one of the important topics in future research on electromagnetic wireless sensing.}
\ifCLASSOPTIONcaptionsoff
  \newpage
\fi

\bibliographystyle{IEEEtran} 
\bibliography{my_reference}

\end{document}